\documentclass[10pt, journal, letterpaper]{IEEEtran}
\usepackage{amsmath}
\usepackage{amsfonts}
\usepackage{blindtext, graphicx,booktabs,verbatim,colortbl,mathtools,amssymb,commath,url,multirow,subcaption,algorithmicx,algpseudocode,algorithm,enumitem,color,tikz}
\usetikzlibrary{arrows}
\usepackage[none]{hyphenat}
\usepackage{amssymb,url}
\usepackage[mathscr]{euscript}
\usepackage{siunitx}
\usepackage{dsfont} 

\usepackage[margin=0.65in]{geometry}

\usepackage{hyperref}

\graphicspath{{images/}}

\usepackage[utf8]{inputenc}
\makeatletter
\newcommand*{\rom}[1]{\expandafter\@slowromancap\romannumeral #1@}
\makeatother

\makeatletter
\def\BState{\State\hskip-\ALG@thistlm}
\makeatother

\usepackage{ulem}
\usepackage[absolute]{textpos}
\ifCLASSINFOpdf
\else
\fi

\DeclareSymbolFont{Xlargesymbols}{OMX}{cmex}{m}{n}
\DeclareMathSymbol{\Xsum}{\mathop}{Xlargesymbols}{80}

\begin{document}

\newpage
\thispagestyle{empty}

\noindent\begin{minipage}{\textwidth}
    {\Huge\textbf{IEEE Copyright Notice}} \\ 

    \vspace{2cm}
    \Large{\copyright 2021 IEEE. Personal use of this material is permitted. Permission from IEEE must be obtained for all other uses, in any current or future media, including reprinting/republishing this material for advertising or promotional purposes, creating new collective works, for resale or redistribution to servers or lists, or reuse of any copyrighted component of this work in other works.}
\end{minipage}

\vspace{2cm} 

\noindent\begin{minipage}{\textwidth}
    
    \LARGE{Accepted for publication in: \textbf{IEEE Transactions on Network and Service Management} \\ \\
    DOI: \textbf{10.1109/TNSM.2021.3086721}}
    
\end{minipage}

\newpage

\title{
Deep Reinforcement Learning-based Content Migration for Edge Content Delivery \\ Networks with Vehicular Nodes}

\title{\textcolor{red}{\small{This paper has been accepted for publication in IEEE Transactions on Network and Service Management (Volume: 18, Issue: 3, September 2021) \\DOI: 10.1109/TNSM.2021.3086721}} \\ [2ex] Deep Reinforcement Learning-based Content Migration for Edge Content Delivery \\ Networks with Vehicular Nodes}

\author{
{Sepideh Malektaji\IEEEauthorrefmark{1},
Amin Ebrahimzadeh\IEEEauthorrefmark{1}, 
Halima Elbiaze\IEEEauthorrefmark{2},
Roch Glitho\IEEEauthorrefmark{1},
and Somayeh Kianpishe\IEEEauthorrefmark{1}}

\IEEEauthorblockA{\IEEEauthorrefmark{1}CIISE, Concordia University, Montr\'eal, Qu\'ebec, Canada}\\
\IEEEauthorblockA{\IEEEauthorrefmark{2}Department of Computer Science, Universit\'{e} du Qu\'{e}bec \`a Montr\'eal, Qu\'ebec, Canada}\\
}
\maketitle

\begin{abstract}
With the explosive demands for data, content delivery networks are facing ever-increasing challenges to meet end-users’ quality-of-experience requirements, especially in terms of delay. Content can be migrated from surrogate servers to local caches closer to end-users to address delay challenges. Unfortunately, these local caches have limited capacities, and when they are fully occupied, it may sometimes be necessary to remove their lower-priority content to accommodate higher-priority content. At other times, it may be necessary to return previously removed content to local caches. Downloading this content from surrogate servers is costly from the perspective of network usage, and potentially detrimental to the end-user QoE in terms of delay. In this paper, we consider an edge content delivery network with vehicular nodes and propose a content migration strategy in which local caches offload their contents to neighboring edge caches whenever feasible, instead of removing their contents when they are fully occupied. This process ensures that more contents remain in the vicinity of end-users. However, selecting which contents to migrate and to which neighboring cache to migrate is a complicated problem. This paper proposes a deep reinforcement learning approach to minimize the cost. Our simulation scenarios realized up to a 70\% reduction of content access delay cost compared to conventional strategies with and without content migration.
\end{abstract}
\IEEEpeerreviewmaketitle
\vspace{-0.80cm}
\section{Introduction}
\IEEEPARstart{T}{he} recent proliferation of mobile devices has provoked an explosive growth in mobile applications and services \cite{vEPC}. These are in turn creating explosive data demands and consequently imposing eruptive traffic loads. Moreover, studies show that the growth of data traffic will be even more drastic in the future. Global mobile data traffic is predicted to increase another seven-fold by 2021 \cite{c2}. It is also predicted that nearly 71\% of this traffic will be distributed by content delivery networks (CDNs) \cite{c2}. A CDN consists of a collection of surrogate servers, placed over multiple locations, with the main objective of delivering content to end-users with the expected quality-of-experience~(QoE). Coping with the explosive growth of data demand while ensuring good QoE is not an easy task for CDN providers. More specifically, the inherent distance between servers and end users induces latency, which remains a roadblock to end-users' QoE. 
Edge computing \cite{c3} can be a promising approach to meet these challenges. In this paradigm, the computing and storage resources located close to end users, known as edge resources, are utilized to decrease the latency. Caching technology is a recognized solution for decreasing content access delay \cite{c4,c5,c6} in wireless networks. Edge computing access nodes such as base stations~(BSs) or road side units~(RSUs) are equipped with small caches, allowing demands to be fulfilled locally with much less delay \cite{c3}. Note that these local edge caches are different than CDN surrogate servers in terms of their caching capacities and distances to end users \cite{c44}. Similar to other caching systems, the precondition to this system's smooth functioning is to know which contents are likely to be demanded in the future and which should be removed. This fundamental requirement has led to many different content placement approaches, such as recently-based (e.g., least recently used (LRU)) \cite{c10} or frequency-based (e.g., least frequently used (LFU)) \cite{c8} policies, or randomized methods \cite{c9} which try to mimic future demands by using the cache history. However, their simple fixed rules can barely adapt to dynamic content access patterns.\par Another important consideration for edge caches is that they are limited in size and often become full, sometimes quite quickly. Eviction policies like LRU and LFU are not applicable for the few contents that may be in an edge cache \cite{c10}, as there are no significant differences in their cached contents' demanded probabilities. Instead, they are generally equally important, and so equally likely to be demanded by local end users in the future. The shortcomings of simple content eviction policies like LRU and LFU could be even more problematic when contents have different priorities. With a system consisting of mobile edge caches, such as vehicular networks, the caching priority of contents could vary widely. As an example, consider a vehicular network consisting of autonomous and non-autonomous vehicles. In this network, the caching priority of certain contents, such as high definition maps (HD maps) \cite{c45}, would generally be higher than that of infotainment contents. Since vehicles' on-board sensors are limited to line-of-sight, autonomous vehicles rely heavily on these maps to plan precisely and maneuver correctly on the road. These machine-readable HD maps model the surface of the road to an accuracy of 10-20 cm and therefore have large volumes \cite{c24}. Since edge caches are limited in size, in the case of full caches, the already placed low-priority contents, such as infotainment contents, will inevitably be deleted to free space for newly demanded high-priority contents. However, evicted contents can be requested again by end users. These evicted contents would need to be retransmitted from cloud servers \cite{c24}. This retransmission is not only time consuming, it is also a waste of network resources. In our example, since the HD maps require periodical updates \cite{c45}, this scenario could happen frequently. The resulting reoccurring delay would have a negative impact on users' quality of experience. 
To address these issues, we propose a novel content migration framework. With this framework, instead of evicting contents completely, edge caches that are full can rely on the available resources of their neighboring caches to migrate selected contents, so that not only can new high-priority contents be cached locally, but previously-cached contents will remain in the vicinity of end users.
This work is an extension of our previously published short paper \cite{c11} in which the proposed genetic-based content placement algorithm considers both the purging probability of contents and the user's priorities. In \cite{c11}, the edge caches were fixed, whereas in this work, we propose a content migration method in a hierarchical CDN consisting of fixed and mobile edge caches.
\par Designing an efficient framework for content migration in a dynamic environment consisting of mobile caches is very challenging; essentially different types of decisions regarding the selection of migrating contents, their sources and destinations must be made almost immediately. Moreover, due to the high speed of mobile caches, the content might not be completely transferred between these edge caches. The mobility of caches, along with the dynamicity of content demands and network conditions, as well as having multi-priority contents, certainly compounds the complexity of the problem. This paper proposes a deep reinforcement learning (DRL)-based framework for content migration in edge CDNs with vehicular nodes. Due to their high mobility and frequently changing topology, integrating a vehicular network with a CDN is indeed challenging \cite{c52}. The key motivation for us is that DRL has been successfully used to tackle similar dynamic and multifactor decision-making problems [21-26]. In this work we design a DRL content migration framework that not only ensures rapid delivery of high-priority contents, but also provides low delay access to existing low-priority contents in a mobile environment. Our proposed framework is based on Q-learning, a model-free RL algorithm \cite{c42}. The high level of complexity in the content migration problem makes the system's state-space large and time-varying, and thus classic Q-Learning approaches are not applicable \cite{c42}. Instead, we rely on recent advances in the DRL field and empower our Q-Learning agent with a specialized deep recurrent neural network (RNN) called LSTM \cite{c21}. The long short-term memory~(LSTM) cells that are used in our RNN ensure the accuracy in the sequential decision-making tasks of content migration. The key contributions of this paper are as follows:

\begin{enumerate}
  \item We introduce a content migration framework for an edge-based CDN consisting of mobile and fixed caches, ensuring that the expensive caching resources at the edge are utilized efficiently. We consider a realistic mobility model that incorporates the characteristic movements, and particularly high speed of mobile caches.
  \item While other works focus primarily on content popularity, our framework not only considers the popularity of contents, but also the different levels of priorities for contents. This aspect makes it applicable to real-world situations where both critical and non-critical contents coexist.
  \item We adopt a deep Q-Learning approach and design a specialized migration agent empowered with LSTM cells that can automatically learn and update its deep Q-Learning network to generate improved migration decisions in a highly dynamic environment. 
\end{enumerate}

\par The remainder of the paper is organized as follows. Section~\ref{sec:relatedwork} presents the motivating scenario, its derived requirements, and a review of the literature. The system model is explained in Section~\ref{sec:system_model}. Section~\ref{sec:optimization} presents the formulation of the content migration optimization problem. Our RL-based approach for solving the content migration problem is presented in Section~\ref{sec:RL} followed by the evaluation of our results in Section~\ref{sec:evaluation}. We conclude the paper in Section~\ref{sec:conclusions}.
\section{Illustrative Use Case, Requirements, and Related Work}\label{sec:relatedwork}
\subsection{Illustrative use case}
Let us consider an edge-based CDN that consists of three layers \cite{c22}. The first layer contains the CDN's content and surrogate servers, which store the original and replicated contents, respectively \cite{c44}. This layer can be considered as the core, as the contents are rooted from this layer. The second, middle layer is comprised of fixed edge nodes such as base stations or roadside units equipped with cache storage \cite{c4}, and the third layer contains mobile edge cache nodes (e.g., autonomous or non-autonomous vehicles with on-board units equipped with limited caching capability \cite{c4}). Figure~\ref{SystemView} illustrates this environment. The contents delivered by this CDN are assumed to have preassigned independent priorities. For instance, infotainment contents have lower priority than the HD maps that are used by autonomous vehicles.  
\par The delay experienced by end users for accessing content varies according to where the content is cached in the CDN's caching network. If the demanded content is already cached in the users' vicinity (for example in a user's own vehicle's onboard cache), the least amount of delay will be imposed for accessing the demanded contents. Fixed-edge caches (e.g., RSUs or BSs), which are in the middle layer of our hierarchical CDN, can also serve their cached contents to users while they are within their coverage range. In this case, a relatively small delay will occur, compared to the case where content is accessed from the CDN's surrogate servers.
\par Next, we consider a concrete scenario, in which end users in an autonomous vehicle are enjoying smooth playback of a popular video that is cached on their vehicle's on-board unit. Since these on-board edge caches are small, the video content occupies a considerable amount of the cache. The autonomous vehicle sends a request for the local HD map. Timely access to these HD maps is very critical, and so after its initial transmission from a CDN server, it should be cached in the closest caching node, which in this case is the vehicle's own on board unit. Moreover, when possible, these HD maps should also be shared with nearby autonomous vehicles. Since both autonomous and non-autonomous cars can coexist along a road, a subset of vehicles,  who are autonomous and located in the vicinity on the same road, are the target caches of a particular HD map.
\par The obvious requirement for caching high-priority contents on target caches is that they should have enough free space. However (as in this scenario), this is not always the case. Especially considering the limited caching capacity of mobile edge caches and the overwhelming volume of both HD maps and infotainment  \cite{c4}, the scenario of inadequate storage space is indeed quite probable. In this case, the low-priority contents (e.g., the infotainment contents), will be dropped. However the dropped contents could be demanded again. Since the retrieval of the content from core servers is time-consuming, the users' QoE, especially in terms of delay, will be affected.
\par We propose an alternative technique, in which instead of being dropped completely, the cached low-priority contents will be migrated to other edge cache nodes in the area, making space for caching the newly arrived high-priority contents while avoiding the need for long delays to retrieve contents from servers. It is important to note that our content migration algorithm considers the case of the high speed movement of mobile caches in which the receiver mobile cache may leave the radio range of a source edge cache before all the bytes of a high- low- priority content could be successfully transferred.

\subsection{Requirements}
Similar to content placement approaches \cite{c16}, a content migration strategy has an impact on the transmission delay and consequently on the end users' QoE. However, a content migration strategy that can be applied on a dynamic network of mobile edge cache nodes should satisfy additional requirements. We summarize these requirements below:
\begin{enumerate}
  \item To ensure a satisfactory QoE, especially the delay requirement for accessing both high- and low-priority contents;
  \item To ensure a low cost for the migration of contents;
  \item To account for the mobility of edge caches as well as end users; and 
  \item To consider the limited capacity of edge caches, especially when they become fully occupied.
\end{enumerate}

\subsection{Related Work}
In the following, we first review the existing research that target the edge content placement and delivery problem in CDNs while considering a content priority scheme. Next, we review the recent DRL-based approaches in the edge caching domain. 

\subsubsection{\textbf{Approaches with content priority schemes}} There are very few works that consider specific content priority schemes in CDNs with edge nodes (and therefore, that satisfy the first requirement). Most of them focus on priority content dissemination rather than caching technology. The work in \cite{c25}, for instance, proposes a priority-based content propagation scheme that accelerates safety content delivery for a set of moving vehicles and provides the forwarding of non-safety contents based on popularity. Similarly, in \cite{c26} an information-centric dissemination protocol for safety information in vehicular ad-hoc networks was proposed. In \cite{c26} a context-aware information dissemination scheme for vehicular networks was introduced, where instead of disseminating all events, only the most relevant ones are disseminated. The authors of \cite{c27} proposed an SDN architecture that uses a data cognitive engine to determine user priority (based on the users' health situation) and allocates edge resources (including edge caching resources) accordingly through a resource-cognitive engine. However, none of the above-mentioned works consider the limited caching capability of edge caches in their solutions, and so none of them propose a strategy for content eviction. Even though both requirements 1 and 3 are satisfied in \cite{c25,c26,c27}, requirements 2 and 4 remain unmet. 
\subsubsection{\textbf{DRL-based approaches for edge content caching}} The use of deep reinforcement learning (DRL) has become quite popular in the networking domain. In a recent work, the authors of \cite{c41} conducted a comprehensive survey on DRL applications for solving a variety of networking problems (e.g., dynamic network access, wireless caching, and data rate control). Specifically, as stated in \cite{c41}, the adoption of DRL for edge caching has received more attention than other networking issues. Zhu et. al. \cite{c12} advocated the use of DRL by examining key challenges in mobile edge caching and then mapping them with unique DRL aspects. The existing edge caching DRL-based approaches can be classified into two categories: ($i$)~works that use DRL for learning specific caching parameters (e.g., content popularity \cite{c42} or cache expiration time), and ($ii$) DRL approaches that target multiple aspects in their caching policy design \cite{c17,c18} (e.g., networking and computation). Falling into the first category, the authors in \cite{c42} propose a DRL-based cache replacement scheme for a single BS, where the content popularity is learned by considering the cache hit rate as the system reward. Similarly, in their recent work \cite{c50}, Yu et. al. propose a federated learning approach to predict content popularity for connected vehicles and provide a mobility-aware cache replacement policy. Many recent works in the vehicular network domain  belong to the second category. For instance, Hu et. al. \cite{c17} proposed an integrated networking, caching, and computing optimization framework for connected vehicles that sets both operational excellence and cost efficiency as objectives. They adopted deep reinforcement learning to overcome the high level of complexity caused by the joint optimization problem. However, they do not suggest a strategy for the case of full edge caches. Similarly, in the recent work of Qiao. et. al. \cite{c51}, DRL is utilised  to solve the joint optimization of content placement and delivery problem in the vehicular networks, formulated as a double time-scale Markov decision process. He. et. al. \cite{c18} proposed a dynamic orchestration framework for communication, caching, and computing resources in a software-defined and virtualized vehicular network. They applied DRL to obtain a close-to-optimal policy for integrated resource allocation. However, no concrete solution for the case of full edge caches is proposed. Considering that resources in edge caches are indeed limited, it is quite probable that these caches become fully occupied. Therefore, having no strategy for these cases is a notable shortcoming of these DRL-based edge caching methods. 
\begin{figure*}[!t]
\centering
   \includegraphics[width=0.65\textwidth]{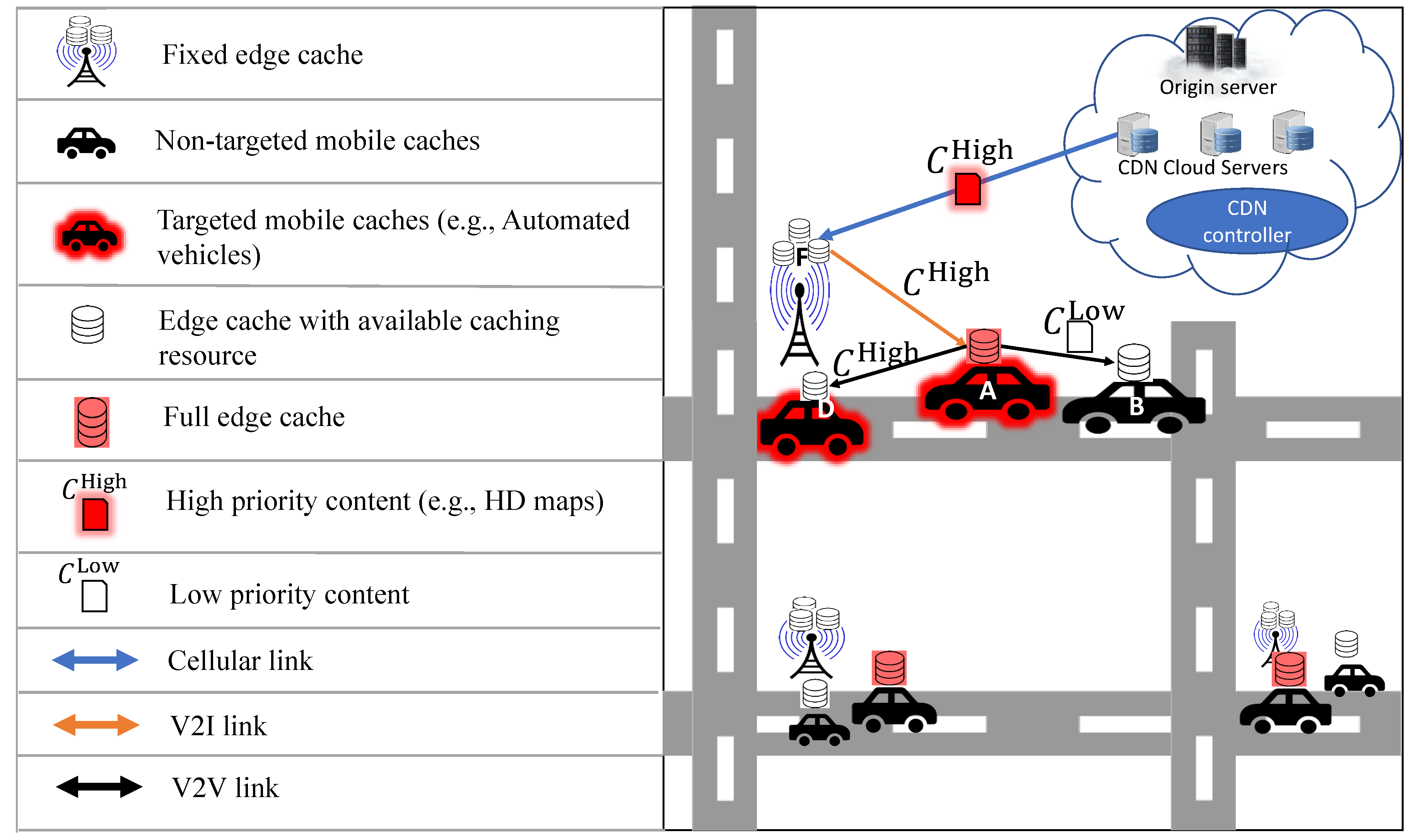}
  \caption{System view and an example of an edge-based CDN with vehicular nodes.}
  \label{SystemView}
 \end{figure*}
\section{System Model}\label{sec:system_model}
Figure \ref{SystemView} illustrates a high-level view of our system model with an example of the problem under study. In Fig.~\ref{SystemView}, certain mobile caches such as autonomous vehicles in an area (referred to as targeted caches) send their requests for high-priority contents to the CDN controller. These high-priority contents could be HD maps and are denoted by $C^{\text{High}}$. Some targeted mobile caches, such as vehicle `A' in Fig.~\ref{SystemView}, may already be fully occupied with low-priority contents denoted by  $C^{\text{Low}}$.  This condition triggers our proposed algorithm to find a desirable solution. The content migration algorithm would free up enough space in target caches to accommodate high-priority contents. However, instead of dropping the low-priority contents, the algorithm provides a new placement for them so that the low-priority contents migrate to nearby available caching resources and thus remain in the vicinity of the target caches. Such a content migration consumes resources both from the host and the destination node. It also consumes scarce bandwidth resources and thus is not cost-free. Our algorithm considers these costs and accordingly proposes a low-cost content migration solution. It also determines a low-cost delivery strategy for high-priority content from edge caches to target caches. 
For example, in Fig.~\ref{SystemView}, the content migration solution could be as follows: Assume a low-priority content $C^{\text{Low}}$ fully occupies vehicle `A''s mobile cache. However, vehicle `A' needs to cache a high-priority content $C^{\text{High}}$. Considering the different migration solution costs, the algorithm could come up with the following strategy: The low-priority content $C^{\text{Low}}$ cached in vehicle `A' should be migrated to neighbor cache `B' via a vehicle-to-vehicle (V2V) link~\cite{c24}. This would free up enough space in the cache of vehicle `A' to store $C^{\text{High}}$. As for the high-priority content delivery strategy, fixed cache `F' can deliver $C^{\text{High}}$ to mobile cache `A' via a vehicle-to-infrastructure (V2I) link~\cite{c24}. The received high-priority content $C^{\text{High}}$ can then be delivered from vehicle `A' to the other targeted caches such as vehicle`D' via a V2V link. This example is valid based on the assumption that vehicle `A' and both vehicles `B' and `D' stay within the coverage ranges of fixed cache `F' and vehicle `A', respectively, long enough for the contents to be transferred successfully. We note that any other migration and delivery strategies could impose a higher cost on the system. In the following, we present the modeling of our considered content migration problem followed by an explanation of the cost calculation for content migration.
\par \textbf{Edge Caches:} We consider a CDN system that consists of mobile and fixed edge caches as well as CDN cloud servers (see Fig.~\ref{SystemView}). Let \( \mathcal{M} = \{e_j\}_{j=1}^{M} \) and \( \mathcal{F} = \{e_f\}_{f=1}^{F} \) denote the sets of mobile and fixed caches with a total number of \text{M} and \text{F} caches, respectively. The total set \( \mathcal{E = M \cup F}\) consists of all the fixed and mobile caches, containing a total of $N=M+F$ edge caches. Each cache \( e_i \in \mathcal{E}\)  has limited caching and processing capacities, denoted by \({L_{stor}(e_i)}\) and \({L_{proc}(e_i)}\), respectively.
\par \textbf{Coverage regions of caches:}
Each cache (fixed or mobile) \( e_i \in \mathcal{E}\) has a circular coverage region with a diameter $\ell_i$. The coverage area of fixed caches is usually larger than that of mobile caches, and thus able to cover a segment of a bidirectional road or an intersection on a given road. We denote the set of fixed caches that have a road intersection in their coverage as \( \mathcal{F}^+\), and those that only cover straight segments of roads as \( \mathcal{F}^-\). Note that the total set of fixed edge caches is  \( \mathcal{F} = \mathcal{F}^+ \cup \mathcal{F}^-\).
\par \textbf{Locations and mobility of caches:} Let $l_i(t)$ be the location of mobile cache \( e_i \in \mathcal{M}\) at time $t$, and $l_f$ be the location of fixed cache \( e_f \in \mathcal{F}\). The velocity of mobile cache \( e_i \in \mathcal{M}\) at time t is denoted by $\mathit{v_i(t)}$. 
For the movement of mobile caches, we adopt a probabilistic model, where mobile caches follow a probabilistic approach in the selection of their direction in a grid-like environment. At each intersection, the mobile cache chooses to keep moving in the same direction or to change direction. The probability of going straight is denoted by $\mu_{\mathscr{S}}$, while taking a left or a right occurs with the probability of $\mu_{\mathsf{L}}$ and $\mu_{\mathscr{R}}$, respectively.
\par \textbf{Contents:} Contents in the considered CDN system have either low or high priority. Let $C^{\text{High}}(t)$ and $C^{\text{Low}}(t)$ denote the high- and low-priority content sets with sizes of $\mathit{\text{size}(C^{\text{High}}(t))}$ and $\mathit{\text{size}(C^{\text{Low}}(t))}$, respectively. $C^{\text{High}}(t)$ and $ C^{\text{Low}}(t)$ contain $\mathit{|C^{\text{High}}(t)|}$ and $\mathit{|C^{\text{Low}}(t)|}$ number of individual contents denoted as $\mathit{c_h^{\text{high}} \in C^{\text{High}}(t)}$ and $\mathit{c_l^{\text{low}} \in C^{\text{Low}}(t)}$, respectively. Following the above-mentioned notations, $\mathit{C(t) = C^{\text{High}}(t) \cup C^{\text{Low}}(t)}$ denotes the total contents at the edge caches at time $t$, where $\mathit{|C(t)|}$ represents the total number of individual contents $\mathit{c_k \in C(t)}$, and $\text{size}(\mathit{C(t)})$ is the total size of the set $\mathit{C(t)}$ in bytes.
\par \textbf{Requests:} Let ${R_{k,i}(t)}$ denote the set of requests for content $\mathit{c_k \in C(t)}$ received by cache ${e_i} \in \mathcal{E}$ at time $t$. $\mathit{|R_{k,i}(t)|}$ also denotes the exact total number of such requests at time $t$. These requests might come from the users within the coverage of ${e_i}$, or they could be the requests of other caches redirected to it. We assume that each request needs $W$ processing units for fulfillment. Therefore, $\mathit{R_{\text{max}}(e_i) = \frac{L_{proc}(e_i)}{W}}$ specifies the maximum number of requests that ${e_i^{\mathcal{E}}}$ can serve simultaneously, given that it has the requested content. 
\par \textbf{Target caches:} In this paper, we label a mobile cache that must locate high-priority content as a `Target Cache'. One example of such mobile caches are autonomous vehicles that need to repeatedly cache an updated version of the HD maps (i.e., high-priority contents) \cite{c24}. Let $\mathit{Q(t)=\{{e_q}\}_{1 \leq q \leq \text{Q}}}$ denote the set of target caches for contents $C^{\text{High}}(t)$. Note that the set of target caches can change over time. If at least one target cache in the set $\mathit{Q(t)}$ does not have enough free storage to accommodate $C^{\text{High}}(t)$, then the content migration strategy should be applied. As a result of applying content migration, a new placement solution for existing low-priority contents will be realized, thereby freeing space for target caches to store high-priority contents. Our proposed algorithm also determines the best delivery strategy for the high-priority contents. Note that as mentioned earlier, we assume a hierarchical structure in CDN caches (see Fig.~\ref{SystemView}) so that the contents first arrive at fixed caches (i.e., RSUs) and from there are distributed to mobile caches. Thus, the delivery strategy for high-priority contents considers the transmission of $C^{\text{High}}(t)$ from fixed caches to target caches.
\par \textbf{Delay:} The average communication delay $D_{i,j}(t,B)$ for transmitting a data of length B (in bytes) from edge cache $\mathit{e_j}$ to $\mathit{e_i}$, where $\mathit{e_i}$ and $\mathit{e_j} \in \mathcal{E}$ at time \textit{t} is estimated as follows
\begin{equation}\label{delay}
  D_{i,j}(t,B) =
  \begin{cases}
  0 & ~\text{if  $i=j$ }\\ 
  \frac{B}{\pounds _{i,j}(t)}+\tau_{i,j} & ~\text{if $i \neq j$ },\\ 
  \end{cases}
\end{equation}
where $\pounds_{i,j}(t)$ and $\tau_{i,j}$ are the  data rate and propagation delay between edge cache $\mathit{e_i}$ to $\mathit{e_j}$ at time \textit{t}, respectively. Further, we model the average communication delay between the edge caches and remote cloud server as a fixed value $d_{\infty}$ (which is dominated by the propagation delay, assuming that the remote cloud server is located hundreds of miles away).
\par \textbf{Power consumption and bandwidth occupation:} Migrating contents from one edge cache to another consumes electrical power, as the source and destination edge caches need to upload and download the content, respectively. We define $g_i$ and $p_i$ as the power consumption cost on edge cache $e_i$ for uploading and downloading one byte of data, respectively. In addition, the network bandwidth will be occupied while migrating the contents. We denote by $\phi$ the bandwidth occupation cost for transmitting one byte of data over one unit of distance in the edge network.


\renewcommand{\arraystretch}{}
\begin{table}[!t]
\normalsize
\centering
\caption{Input parameters and variables.}\label{variables}
\label{table}
\resizebox{\columnwidth}{!}{%
\begin{tabular}{|p{1.3cm}|p{11cm}|}
\hline
\rowcolor{lightgray}\multicolumn{2}{|c|}{\textbf{Network Parameters}}                                                                               \\ 
\hline
$\mathcal{M}$                 & Set of $M$ number of mobile edge caches                                                                                                         \\ \hline
$\mathcal{F}$                 & Set of $F$ number of fixed edge caches                                                                                                                  \\ \hline
$\mathcal{E}$                 & Set of $N=M+F$ number of edge caches, $\mathcal{E}=\mathcal{M} \cup \mathcal{F}$                                                                                                       \\ \hline
 $\mathcal{F}^+$            &Set of fixed caches covering a road intersection          \\ \hline
$\mathcal{F}^-$        & Set of fixed caches covering a straight road segment                     \\ \hline
$g_i$ & Power consumption cost of $e_i$ for uploading one byte
   \\ \hline
$p_i$ & Power consumption cost of $e_i$ for downloading one byte
   \\ \hline
$\phi$ & Bandwidth cost of one byte transmitted for one hop
\\ \hline
$d_{\infty}$  &   Delay of transmitting a byte from cloud servers to edge caches
\\\hline
$\mathit{C(t)}$ & Set of all contents at time $t$ including low- and high-priority contents, $\mathit{C(t) = C^{\text{High}}(t) \cup C^{\text{Low}}(t)}$ 
\\\hline
$\mathcal{Q}(t)$         &  Set of target caches for $C^{High}(t)$              \\\hline
$\mathcal{D}(t)$         &  Set of dominating caches at time t              
\\\hline
$\mathit{v}_i(t)$        & The instantaneous speed of mobile cache $e_i$ at time $t$                 \\ \hline
$\ell_i$ & Diameter of circular coverage of edge cache $e_i$    
      \\ \hline
$e_i$         & Edge cache $i$ ($e_i  \in \mathcal{E}$)                                         \\ \hline
\hspace*{-3mm} $R_{\text{max}}(e_i)$        & Maximum number of requests $e_i$ can serve simultaneously  
\\ \hline 
$\varrho_{i,j}(t)$                 & Sojourn time of edge cache $e_j$  in coverage of edge cache $e_i$                  \\ \hline 
$r_{i,j}$                 & Length of road path in coverage of  $e_i$  traversed by $e_j$ when \(e_i,e_j \in \mathcal{M}\)
\\ \hline
$r^{\mathscr{S}}_{f,u}$                 & The length of straight road path within the coverage of fixed edge cache $e_f$ traversed by $e_u \in  \mathcal{M}$   \\ \hline   
$r^{\mathsf{L}}_{f,u}$                 & The length of left road path within the coverage of fixed edge cache $e_f$ traversed by $e_u \in  \mathcal{M}$     \\ \hline 
$r^{\mathscr{R}}_{f,u}$                 & The length of right road path within the coverage of fixed edge cache $e_f$ traversed by $e_u \in  \mathcal{M}$\\ \hline
$\mu_{\mathscr{S}}$              & The probability that mobile caches follow a straight road.
\\ \hline
$\mu_{\mathscr{R}}$              & The probability that mobile caches take a right turn.
\\ \hline
$\mu_{\mathsf{L}}$              & The probability that mobile caches take a left turn.
\\ \hline
$L_{stor}(e_i)$ & Caching capacity of $e_i \in \mathcal{E}$
\\ \hline
$L_{proc}(e_i)$ & Processing capacity of $e_i\in \mathcal{E}$
\\ \hline
${l_i(t)}$         &  Location of edge cache $e_i \in \mathcal{E}$            
\\\hline

$\gamma$ & Delay (1ms) cost to access one byte of low-priority content
\\ \hline
$\psi$ & Delay (1ms) cost to access one byte of high-priority content
\\ \hline
$\mathit{{R}_{k,i}(t)}$ & Set of requests for content $c_k \in C(t)$ received by $e_i$ at time $t$ 
\\ \hline
$D_{i,j}(t,B)$  &   Delay of transmitting B bytes from $e_i$  to $e_j$   at time $t$
 \\ \hline
 $\pounds_{i,j}$ & average data rate between edge cache $e_i$  and $e_j$.
 \\ \hline 
 $\tau_{i,j}$ & Propagation delay between edge cache $e_i$  and $e_j$.
 \\ \hline
$\lambda_{c_k}$  &   Average request rate for content $c_k$
\\ \hline                                      
$\flat_{i,j,l}(t)$                 & Number of downloaded bytes of $\mathit{c_l^{\text{low}}}$ from $e_i$  to $e_j$. $e_i,e_j \in \mathcal{E}$                         \\ \hline  
$\vartheta_{u,f,h}(t)$  &Number of downloaded bytes of $\mathit{c_h^{\text{high}}}$ from $e_f \in \mathcal{F}$  to $e_u \in \mathcal{D}(t)$                  \\ \hline
$\Im_{u,f,h}$                 & Delay of transmitting $\mathit{c_h^{\text{high}}}$ from $e_f \in \mathcal{F}$  to $e_u \in \mathcal{D}(t)$                          \\ \hline   

\end{tabular}
}

\end{table}


\section{OPTIMIZATION MODEL FOR CONTENT MIGRATION}\label{sec:optimization}

Our objective is to decide how to place high-priority contents in a subset of fixed caches to deliver them to target caches. This task includes determining how to place low-priority contents in mobile caches such that the aggregated cost of content migration and the content access delay are minimized, while ensuring the target caches have enough free space to accommodate the high-priority contents. In this section, we introduce the set of input parameters and decision variables considered in our formulation, and then explain our objective function and constraints. Table~\ref{table} delineates some of the important inputs and variables used in our formulation. The optimization decision variables are defined as follows:

\begin{itemize}
    \item $y_{i,l}(t)$: A binary decision variable which is 1 when low-priority content $c_l^{\text{low}}$  is in the edge cache $\mathit{e_i}$ at time t (otherwise it is 0).
    \item $x_{f,h}(t)$: A binary decision variable which is 1 when high-priority content $\mathit{c_h^{\text{high}}}$ is in the fixed edge cache $\mathit{e_f}$ at time \textit{t} (otherwise it is 0).
    \item $z_{i,j,k}(t)$: An integer decision variable between 0 and $R_{\text{max}}(e_i)$. This variable identifies the number of requests for content $\mathit{c_k} \in C(t)$ redirected from  $\mathit{e_j}$ to $\mathit{e_i}$ at time $t$. Note that for the special cases where $j=i$, the parameter $z_{i,j,k}(t)$ represents the number of requests for content $\mathit{c_k}$, which are received and directly processed by $\mathit{e_j}$ itself.
\end{itemize}

\subsection{Content Migration Cost}
Content migration cost at time \textit{t} consists of three partial cost components, $C_1(t)$, $C_2(t)$, and $C_3(t)$. The first partial cost, $C_1(t)$, is the cumulative cost of the power consumption associated with uploading contents from edge caches, given by
\begin{equation}\label{C1}
  C_{1}(t) = \Xsum_{i=1}^N \Xsum_{l=1}^{|C^{\text{Low}}(t)|} \Lambda_{l,i} \cdot
\left({y_{i,l}(t-1)-y_{i,l}(t)}\right)^+, 
\end{equation}
where
\begin{equation}
  {(A-B)}^+=\begin{cases}
    1, & \text{if $A > B$}\\
    0, & \text{otherwise}.
  \end{cases}
\end{equation}
For each content $\mathit{c_l^{\text{low}}}$, $({y_{i,l}(t-1)-y_{i,l}(t)})^+$ is equal to 1 when content $\mathit{c_l^{\text{low}}}$ is uploaded from $\mathit{e_i}$. In this case, the non-negligible cost of $\Lambda_{l,i} = \text{size}(\mathit{c_l^{\text{low}}})\cdot g_i$ will be imposed on the system due to content uploading. 

Similarly, the cumulative cost of power consumption $C_2(t)$ associated with downloading contents from edge caches at time $t$ is given by
\begin{equation}\label{C2}
  \begin{split}
  C_{2}(t) = \Xsum_{i=1}^N \Xsum_{l=1}^{|C^{\text{Low}}(t)|} V_{l,i}\cdot
\left({y_{i,l}(t)-y_{i,l}(t-1)}\right)^+.
\end{split}
\end{equation}
In this case, the non-negligible cost of $V_{l,i} = \text{size}(\mathit{c_l^{\text{low}}})\cdot p_i$ will be imposed on the system due to content downloading.

The cost $C_3(t)$ of bandwidth occupation involved in migrating contents between edge caches is given by
%
%
%
\begin{equation}
\begin{aligned}
C_{3}(t) =&\Xsum_{i,j=1 \atop
i \ne j}^N \Xsum_{l=1}^{|C^{\text{Low}}(t)|}\Delta_l.\mid l_i(t)- l_j(t)\mid \cdot \\
& \left[{y_{i,l}(t-1)-y_{i,l}(t)}\right]^+ \cdot \left[{y_{j,l}(t)-y_{j,l}(t-1)}\right]^+,
\label{C3}
\end{aligned}
\end{equation}
where the term$\left[{y_{i,l}(t-1)-y_{i,l}(t)}\right]^+ \cdot \left[{y_{j,l}(t)-y_{j,l}(t-1)}\right]^+$ becomes non-zero only when content $\mathit{c_l^{\text{low}}}$
is uploaded from $\mathit{e_i}$ and downloaded into $\mathit{e_j}$. Moreover, letting $\phi$ be the bandwidth occupation cost for transmitting one byte over a unit of distance, $\Delta_l =\text{size}(\mathit{c_l^{\text{low}}})\cdot\phi$ will be equal to the bandwidth occupation cost for transferring $\mathit{c_l^{\text{low}}}$ over a unit of distance. As (\ref{C3}) suggests, the associated bandwidth occupation cost can be calculated by multiplying $\Delta_l$ by the distance $\mathit{\mid l_i(t)- l_j(t)\mid}$ between source and destination. 
The content migration cost $C_{M}$ is then obtained by summing the three partial costs $C_1(t)$, $C_2(t)$, and $C_3(t)$ accumulated over the observation time period $[{t_0}, {t_k}]$ as follows:
\begin{equation}\label{Cmig}
    C_{M} = \int_{t_0}^{t_k} [C_1(t) + C_2(t) + C_3(t)] \cdot \textit{dt}.
\end{equation}

\subsection{Delay cost of low-priority contents}
To calculate the delay cost of low-priority contents, we assume that the content popularity follows a Zipf distribution~\cite{c18,c48} with $\alpha$ being the Zipf slope ($0 < \alpha < 1$). Assuming $\mathit{c_l^{low}}$ is the $l$'th most popular content, the probability of content $\mathit{c_l^{low}}$ being requested is $\frac{1}{\rho.l^\alpha}$, where $\rho = \sum^{\mathit{\text{size}(C^{\text{Low}}(t))}}_{l=1}\frac{1}{l}$. With the assumption that the low-priority content requests follow a Poisson process with parameter $\beta$~\cite{c18,c48}, the average request rate $\lambda _{l}$ of content $\mathit{c_l^{low}}$ can be calculated by:
\begin{equation}\label{lambdaa}
   \lambda _{l} = \frac{\beta}{\rho.l^\alpha}  
\end{equation}

\begin{figure*}[!t]
\centering
   \includegraphics[width=0.75\textwidth]{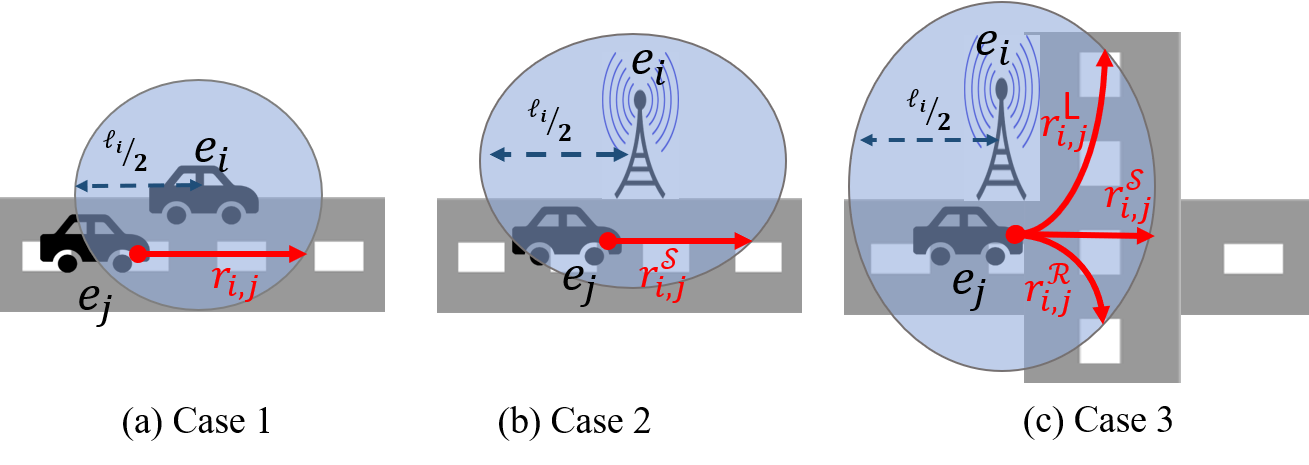}
  \caption{Illustration of the three different cases for calculating the sojourn time $\varrho_{i,j}(t)$.}
  \label{CasesFig}
 \end{figure*}

 The delay cost $C_{A}$ of accessing low-priority contents is given by
\begin{equation}\label{Caccess}
 \begin{split}
  \mathit{C_{A}} = \int_{t_0}^{t_k} 
 \mathit{\Xsum_{i,j=1 \atop i \ne j}^N \Xsum_{l=1}^{|C^{Low}(t)|}} \lambda_{l}.\gamma.
   [D_{i,j}(t,\flat_{i,j,l})\cdot z_{i,j,l}(t) \cdot y_{i,l}(t) + \\
   d_{\infty} \cdot (|R_{l,i}(t)| - z_{i,j,l}(t) \cdot y_{i,l}(t))] \cdot dt,
\end{split}
\end{equation}
where the term $z_{i,j,l}(t) \cdot y_{i,l}(t)$ counts the number of requests for content $\mathit{c_l^{low}}$ that are sent to edge cache $\mathit{e_i}$, and $\gamma$ is the cost of a unit delay for accessing one byte of low-priority content. 
However, some of the requests may not be fulfilled successfully, which occurs when the receiving edge node leaves the radio coverage of the transmitter node before the whole content has been transmitted. To compute the successfully transmitted bytes of $\mathit{c_l^{low}}$, we first obtain the sojourn time $\varrho_{i,j}(t)$ of $\mathit{e_j}$ in the coverage area of $\mathit{e_i}$. Depending on the type of the caches (i.e., fixed or mobile) and their coverage area, we may deal with one of the following three cases to calculate the sojourn time $\varrho_{i,j}(t)$, as shown in Fig. \ref{CasesFig}.

\textbf{Case 1:} The two caches $\mathit{e_j}$ and $\mathit{e_i}$ are both mobile, i.e., $\mathit{e_i},\mathit{e_j} \in \mathcal{M}$ (see Fig~\ref{CasesFig}a). In this case, the sojourn time $\varrho_{i,j}(t)$ is given by:

\begin{subequations}

\begin{equation}\label{Sojourn-a}
\begin{split}
\varrho_{i,j}(t) = [\frac{\ell_i}{2}-|l_i(t)-l_j(t)|]^+ \cdot \frac{r_{i,j}}{|v_i(t)-v_j(t)|}, & \hspace{+4cm} 
\end{split}
\tag{8a}
\end{equation}
where the term $[\frac{\ell_i}{2}-|l_i(t)-l_j(t)|]^+$ is equal to 1 only if $e_j$ resides within the coverage of $e_i$ at time $t$; otherwise, it is zero. In Eq.~(\ref{Sojourn-a}) $r_{i,j}$ is the length of road path within the coverage of  $e_i$  traversed by $e_j$, and $|v_i(t)-v_j(t)|$ is the relative speed of $e_j$ with respect to $e_i$.
\textbf{Case 2:} In this case, $e_j$ is mobile, whereas $e_i$ is a fixed cache covering a straight road segment; i.e., $\mathit{e_j} \in \mathcal{M}$ and $\mathit{e_i} \in \mathcal{F^-}$ (see Fig.~\ref{CasesFig}b). We can then calculate $\varrho_{i,j}(t)$ as follows:
\begin{equation}\label{Sojourn-b}
\begin{split}
\varrho_{i,j}(t) = [\frac{\ell_i}{2}-|l_i-l_j(t)|]^+\cdot \frac{r^{\mathscr{S}}_{i,j}}{\mathit{v_j(t)}}, & \hspace{+4cm} 
\end{split}
\tag{8b}
\end{equation}
where $r^{\mathscr{S}}_{i,j}$ is the length of straight road path within the coverage of fixed edge cache $e_i$ traversed by $e_j$.

\textbf{Case 3:} In the third case, $e_j$ is mobile and $e_i$ is a fixed cache covering a road intersection; i.e., $\mathit{e_j} \in \mathcal{M}$ and $\mathit{e_i} \in \mathcal{F^+}$  (see Fig.~\ref{CasesFig}c). We can then calculate $\varrho_{i,j}(t)$ as follows:
\begin{equation}\label{Sojourn-c}
\begin{split}
\varrho_{i,j}(t) = [\frac{\ell_i}{2}-|l_i-l_j(t)|]^+ \cdot \frac{\mu_{\mathscr{S}}\cdot{r^{\mathscr{S}}_{i,j}}+\mu_{\mathsf{L}}\cdot{r^{\mathsf{L}}_{i,j}}+\mu_{\mathscr{R}}\cdot{r^{\mathscr{R}}_{i,j}}}{\mathit{v_j}(t)}, & \hspace{+4cm} 
\end{split}
\tag{8c}
\end{equation}
\end{subequations}
where $r^{\mathscr{S}}_{i,j}$ is the length of straight road path within the coverage of fixed edge cache $e_i$ traversed by $e_j$, while $\mu_{\mathscr{S}}$ is the probability that mobile caches follow a straight road. $r^{\mathsf{L}}_{i,j}$ is the length of left road path within the coverage of fixed edge cache $e_i$ traversed by $e_j$, while $\mu_{\mathsf{L}}$ is the probability that a mobile cache takes a left turn. Further, $r^{\mathscr{R}}_{i,j}$ is the length of right road path within the coverage of fixed edge cache $e_i$ traversed by $e_j$, while $\mu_{\mathscr{R}}$ is the probability that a mobile cache takes a right turn. We note that upon facing an intersection along its path, a mobile cache follows the straight road or takes a left or right turn with probabilities $\mu_{\mathscr{S}}, \mu_{\mathsf{L}}$, and $\mu_{\mathscr{R}}$, respectively.

By setting $D_{i,j}(t,B) = \varrho_{j,i}$ in Eq.~(\ref{delay}), the number of bytes that can be transferred from $\mathit{e_i}$ to $\mathit{e_j}$ can be computed (i.e., the term $[{\varrho_{j,i}(t)}-\tau_{i,j}]\cdot\pounds_{i,j}(t)$). Specifically, the number of successfully transmitted bytes of $\mathit{c_l^{\text{low}}}$ from edge cache $\mathit{e_i}$ to $\mathit{e_j}$ can be obtained as follows: 
\begin{equation}\label{va}
  \flat_{i,j,l}(t) = \min\bigl\{y_{i,l}(t) \cdot ([{\varrho_{j,i}(t)}-\tau_{i,j}]\cdot\pounds_{i,j}) ,\text{size}(c_l^{\text{low}})\bigr\}.  
\end{equation}
It should be noted that in (\ref{Caccess}) any requests for the remaining bytes of $\mathit{c_l^{\text{low}}}$ that can not be served from any edge caches are assumed to be redirected to cloud servers for fulfillment. 

\subsection{Delay cost of high-priority contents}
Target cache $\mathit{e_q} \in \mathcal{Q}(t)$ can download high-priority contents $C^{\text{High}(t)}$ either directly from fixed caches or from other target caches that have already received high-priority contents. However, if $\mathit{e_q}$ cannot receive $C^{\text{High}(t)}$ from any other edge caches due to, for instance, their isolated location or high speed, it should download all or the remaining parts of those contents from CDN cloud servers. %
We define the so-called dominating cache $\mathcal{D}(t)$ as a subset of target caches that can transmit the contents to the rest of the target caches with at most $\sigma$ hops. Note that $\mathcal{D}(t) \subseteq \mathcal{Q}(t)$ and that it is identified by means of graph theory. Let the contact graph $\mathit{G(t)= ( e_q |\forall q \in Q(t) , \mathit{E}(t))}$ be the representation of the target caches' topology at time $t$, where $\mathit{E}(t)$ is the set of edges showing the connectivity among the target caches. In this regard, edge $\xi_{q,p}(t)\in \mathit{E}(t)$ exists if and only if the target caches $\mathit{e_q}$ and $\mathit{e_p}$ are in the transmission range of each other at time $t$. The set of dominating nodes in the contact graph can be identified in polynomial time by the algorithm proposed in~\cite{c29}. 
Depending on its path, dominating cache $\mathit{e_u} \in \mathcal{D}(t)$ receives a high-priority content $\mathit{c_h^{\text{high}} \in C^{\text{High}}(t)}$ in a continuous manner while switching from one fixed cache range to another. 
The number of bytes in content $\mathit{c_h^{\text{high}}}$ downloaded from $\mathit{e_f}$ varies according to the amount of time $\mathit{e_u} \in \mathcal{D}(t)$ spends in the coverage area of fixed cache $\mathit{e_f}$. Recall that $\varrho_{u,f}(t)$ (estimated by Eqs.~(\ref{Sojourn-a}), (\ref{Sojourn-b}), and (\ref{Sojourn-c}) above) is the duration time that $\mathit{e_u} \in \mathcal{D}(t)$ remains within the coverage range of fixed cache $\mathit{e_f}$. Using $\varrho_{u,f}(t)$, the number of successfully transmitted bytes of content $\mathit{c_h^{\text{high}}}$ from fixed cache $\mathit{e_f}$ to target cache $\mathit{e_u} \in \mathcal{D}(t)$ can be calculated as follows:  
\begin{equation}\label{vartheta}
  \vartheta_{u,f,h}(t) = \min\bigl\{x_{f,h}(t) \cdot ([{\varrho_{u,f}(t)}-\tau_{f,u}]\cdot\pounds_{f,u}),\text{size}(c_h^{high})\bigr\}.  
\end{equation}
Considering (\ref{delay}), the delay of downloading high-priority content $\mathit{c_h^{\text{high}}}$ from fixed edge cache $\mathit{e_f}$ to $\mathit{e_u} \in \mathcal{D}(t)$ can be obtained as:
\begin{equation}\label{im}
  \Im_{u,f,h}(t) = D_{u,f}(t,\vartheta_{u,f,h}(t)). 
\end{equation}
We note that when $\mathit{e_u} \in \mathcal{D}(t)$ is out of the range of any fixed cache, the request should be redirected to the cloud and the remaining portions of the high-priority content downloaded from CDN cloud servers. The term $(|R_{h,u}(t)| - z_{f,u,h}(t) \cdot x_{f,h}(t))$ in Eq.~(\ref{Cdeliv}) computes the number of such requests.
The high-priority content delivery cost for target caches can then be obtained as follows:
\begin{equation}
\begin{aligned}
 &C_{D} = \int_{t_0}^{t_k} (\Xsum_{u|e_u^{DC} \in D(t)}\Xsum_{h=1}^{\mathit{|C^{\text{High}}(t)|}}
(  \Xsum_{f=1}^F \\
  & [  \Im_{u,f,h}(t) + [|R_{h,f}(t)| - z_{f,u,h}(t) \cdot x_{f,h}(t)] \cdot d_{\infty}] \cdot \Psi) \cdot dt,
\label{Cdeliv}
\end{aligned}
\end{equation}
where $\Psi$ is the delay cost per second of downloading one byte of high-priority content.

\subsection{Objective Function and Constraints}\label{sec:objective_function}
The objective is to minimize the total cost as an aggregation of content migration cost and content delay cost in a given CDN system. Let $w_{M}$, $w_{A}$, and $w_{D}$ denote the weights of the costs $C_{M}$, $C_{A}$, and $C_{D}$, respectively. The objective function $\Phi$ over the observation time period $[{t_0}, {t_k}]$ is then given by
\begin{center}
\begin{equation}\label{Ctot}
 \min \: \: \Phi =
   w_{M} \cdot C_{M}+ w_{A} \cdot C_{A}+ w_{D} \cdot C_{D},
\end{equation}
\end{center}
subject to the following constraints: 
%
%
%
\begin{subequations}
\begin{equation}\label{cst1}
\begin{split}
\Xsum_{l=1}^{\mathit{|C^{\text{Low}}(t)|}} \Xsum_{j=1}^M z_{i,j,l}(t) \cdot y_{i,l}(t) \leq &\mathit{R_{\text{max}}(e_i}),
\\
&\forall t_0 \leq t \leq t_{k}, \forall 1\leq i \leq M,
\end{split}
\end{equation}
%
\begin{equation}\label{cst2}
\begin{split}
&\Xsum_{u | \mathit{e_u \in \mathcal{D}(t)}} \Xsum_{h=1}^{ \mathit{|C^{\text{High}}(t)|}}
([ \frac{\ell_f}{2}-(\mathit{l_u(t)-l_f})]^+ \cdot x_{f,h}(t))+ \\
&\Xsum_{l=1}^{\mathit{|C^{\text{Low}}(t)|}} \Xsum_{j=1}^{M} z_{f,j,l}(t) \cdot y_{f,l}(t) \leq \mathit{R_{\text{max}}(e_f}),\\
& \qquad \qquad \qquad \qquad \qquad \qquad \qquad \forall t_0 \leq t \leq t_{k}, \forall 1\leq f \leq F,
 \end{split}
\end{equation}
%
\begin{equation}\label{cst3}
\begin{split}
\Xsum_{l=1}^{\mathit{|C^{\text{Low}}(t)|}} y_{v,l}(t) \cdot \mathit{\text{size}(c_l^{\text{Low}}}) &\leq \mathit{L_{stor}(e_v)},\\
&\forall t_0 \leq t \leq t_{k}, \forall \mathit{e_v} \in \mathit{\{\mathcal{M} - \mathcal{Q}(t)\}},
\end{split}
\end{equation}
\begin{equation}\label{cst4}
\begin{split}
\Xsum_{l=1}^{\mathit{|C^{\text{Low}}(t)|}} y_{q,l}(t) \cdot \mathit{\text{size}(c_l^{\text{Low}})} + \mathit{\text{size}(C^{\text{High}}(t))}\leq \mathit{L_{stor}(e_q)},\\
\end{split}
\end{equation}
\begin{equation}\label{cst5}
\begin{split}
\sum_{l=1}^{\mathit{|C^{\text{Low}}(t)|}} y_{f,l}(t) \cdot \mathit{\text{size}(c_l^{\text{Low}})} + 
\Xsum_{h=1}^ {\mathit{|C^{\text{High}}(t)|}} x_{f,h}(t) \cdot \mathit{\text{size}(c_h^{\text{high}})},\\
  \leq \mathit{L_{stor}(e_f)}\\
\forall t_0 \leq t \leq t_{k}, \forall f | \mathit{e_f \in \mathcal{F}}.
\end{split}
\end{equation}
\end{subequations}
%
Constraint~(\ref{cst1}) ensures that the maximum number of content requests that can be served simultaneously from mobile edge cache $\mathit{e_i} \in \mathcal{M}$ is not exceeded. Similarly, constraint~(\ref{cst2}) indicates a set of constraints on the number of requests for contents that can be handled simultaneously by fixed cache $\mathit{e_f} \in \mathcal{F}$. Note that $[\frac{\ell_f}{2}-(l_u(t)-l_f)]^+$ computes the number of dominating caches covered by $\mathit{e_f}$ at time $t$. Therefore, the number of requests for downloading high-priority contents are calculated by the first term in constraint~(\ref{cst2}), while the second term computes the number of requests for low-priority contents. Constraint~(\ref{cst3}) represents the capacity constraints for non-target caches $e_v\in \{ \mathcal{M} - \mathcal{Q}(t)\}$. Similarly, the capacity constraints of target mobile caches are specified by constraint~(\ref{cst4}). Note that each target mobile cache, in addition to the contents already cached in it, should also have space for high-priority contents; this is ensured by constraint~(\ref{cst4}). Finally, constraint~(\ref{cst5}) represents the capacity constraints on fixed caches.

\subsection{Problem Analysis}
By considering the contents as items and caches as bins, the problem defined in Section~\ref{sec:objective_function} can be mapped to the constrained dynamic bin packing problem, which belongs to the class of NP-hard problems~\cite{c30}. This category of problems are usually solved using heuristics or meta heuristics~\cite{c31}. However, these methods are not suitable for our problem because the content migration problem defined above in Section~\ref{sec:objective_function} includes dynamic parameters, such as the location of mobile caches and the number of newly arrived high-priority contents. In fact, the values of such parameters can change rapidly in our content migration problem. In addition, as discussed in \cite{c46}, evolutionary optimization algorithms and (meta)heuristics usually need to be reperformed each time the parameters change, and they cannot automatically adjust their solution in accordance with these changes. 

In light of the above discussion, we aim to utilize an RL-based content migration agent that can automatically capture and update the dynamic statistics of parameter values, and to use these dynamics to quickly adjust its content migration decisions. To elaborate on the design of such an agent, we first present the problem in the form of a Markov Decision Process~(MDP)~\cite{c30}. In the following, we demonstrate that our defined problem preserves the memory-less property (also known as the Markov property)~\cite{c30}. In Section \ref{sec:RL}, we formulate and analyze our RL-based content migration algorithm, taking into account the main MDP components,including the state set, action set, and reward function.
\par In (\ref{Ctot}), the three partial costs $C_{M}$, $C_{A}$, $C_{D}$ can be represented as the accumulation costs over the complete set of observation time slices, indexed by $\tau = 0,..., k$. Let $|T|$ be the length of each time slice and $t_0$ be the first time slice. The weighted cost $\Phi$ in Eq.~(\ref{Ctot}) can be rewritten as: 
\begin{equation}
\begin{aligned}
 \Phi = \sum_{\tau = 0}^{k} \int_{t_0+\tau|T|}^{t_0+(\tau+1)|T|}(w_{M} \cdot C_{M}(\tau)&+ w_{A} \cdot C_{A}(\tau)\\&+ w_{D} \cdot C_{D}(\tau))\cdot dt
\label{wp}
\end{aligned}
\end{equation}
where $C_{M}(\tau)$, $C_{A}(\tau)$, and $C_{D}(\tau)$ are the migration cost, the delay cost of low-priority contents, and the delay cost of high-priority contents, respectively, during time slice $t_\tau$.
We defined $\Phi(\tau)$ as the aggregation of the total costs from time slot $t_0$ to $t_\tau$. We note that $\Phi(\tau)$ can be written as follows: 
\begin{equation}
\begin{aligned}
  \Phi(\tau + 1 ) = \Phi(\tau) + w_{M}\cdot C_{M}(\tau)+ w_{A} \cdot C_{A}(\tau)+ w_{D} \cdot C_{D}(\tau),
\label{wptau}
\end{aligned}
\end{equation}
where the future value $\Phi(\tau+1)$ of the weighted cost function depends only on the current parameters' values and not on their values in previous time slices. This indicates that $\Phi$ has the memory-less property~\cite{c31} and our defined content migration problem can therefore be presented in an MDP, which can be solved using a reinforcement learning approach~\cite{c31}.

\section{RL-based Content Migration}\label{sec:RL}
We define the main components of the MDP in our content migration problem using the formulations presented in Section~\ref{sec:optimization} and show how we utilize this format to develop our RL-based content migration solution.

\par \subsection{System States}
The state of the system at time $t$ should represent ($i$)~the placement of both low- and high-priority contents on the edge caches at that time and ($ii$)~the content delivery state of the system. The delivery state of the system is defined as the participation level of each edge cache in the delivery of requested contents. This participation level is quantified by the number of redirection requests that  each edge cache performs at time $t$. To formally present the system states set, we define $\mathbb{Y}^{c_l}(t)$, $\mathbb{X}^{c_h}(t)$ and $\mathbb{Z}^{c_l}(t)$ as the realization sets of random variables $y_{i,l}(t)$, $x_{f,h}(t)$ and $z_{i,j,l}(t)$ at time $t$, respectively. We note that $\mathbb{Y}^{c_l}(t)$, $\mathbb{X}^{c_h}(t)$ and $\mathbb{Z}^{c_l}(t)$ are given by 
\begin{equation}\label{Y}
     \mathbb{Y}^{c_l}(t) = [Y_{1,l}(t),Y_{2,l}(t)...,Y_{N,l}(t)],
\end{equation}
\begin{equation}\label{X}
     \mathbb{X}^{c_h}(t) = [X_{1,h}(t),X_{2,h}(t)...,X_{F,h}(t)],
\end{equation}
and
\begin{equation}\label{Re}
     \mathbb{Z}^{c_l}(t) = \begin{bmatrix}
Z_{1,1,l}(t)  & Z_{1,2,l}(t)&\cdots & Z_{1,N,l}(t) \\
\vdots & \vdots & \ddots & \vdots\\
Z_{N,1,l}(t) & Z_{N,2,l}(t)&\cdots &Z_{N,N,l}(t)
\end{bmatrix},
\end{equation}
where $Y_{i,l}(t)$, $X_{f,h}(t)$ and $Z_{i,j,l}(t)$ are the exact values of random variables $y_{i,l}(t)$, $x_{f,h}(t)$, and $z_{i,j,l}(t)$ at time $t$, respectively. We then encapsulate $\mathbb{Y}^{c_l}(t)$, $\mathbb{X}^{c_h}(t)$ and $\mathbb{Z}^{c_l}(t)$ in the vector $\chi^{c_h,c_l}(t)$ given by
\begin{equation}\label{chi}
    \chi^{c_h,c_l}(t) = [\mathbb{Y}^{c_l}(t),\mathbb{X}^{c_h}(t),\mathbb{Z}^{c_l}(t)].
\end{equation}
Finally, the state ${s}_{t}$ of the system at time $t$ can be calculated as follows: 
\begin{equation}\label{S}
      {s}_{t}= [\cup\chi^{c_h,c_l}(t)|c_h \in C^{\text{High}}(t), c_l \in C^{\text{Low}}(t)].
\end{equation}

\subsection{System Actions}
The agent can take action by migrating, caching or dropping the contents from edge caches or by redirecting requests between them. To better explain these possible actions, we divide them into three types. The first type of action is to migrate, cache or drop the low-priority content $c_l \in C^{low}(t)$ in edge caches (both fixed and mobile caches). We refer to this action type as {\fontfamily{qcr}\selectfont Act.Type1}. This type of action will cause changes in the values of $\mathbb{Y}^{c_l}(t+1)$ with respect to $\mathbb{Y}^{c_l}(t)$. To represent this type, a binary vector $a_{c_l}^\mathbb{Y}(t)$ of size $N$ is used, where $N$ is the total number of edge caches. A value of 1 for the $i$-th element of vector $a_{c_l}^\mathbb{Y}(t)$ indicates a zero to one or vice versa change in $i$-th element of $\mathbb{Y}^{c_l}(t)$ (i.e. $Y_{i,l}(t)$'s value ), whereas a value of 0 would indicate no change in the value of $Y_{i,l}(t)$. The second type of action, {\fontfamily{qcr}\selectfont Act.Type2}, is the caching (or dropping) of the high-priority content $c_h \in C^{high}(t)$ on (from) fixed caches for their later delivery to targeted caches. The effect on the values of $\mathbb{X}^{c_h}(t)$ will be similar to that of {\fontfamily{qcr}\selectfont Act.Type1}, and we represent it by binary vector $a_{c_h}^\mathbb{X}(t)$ of size $F$, where $F$ is the number of fixed edge caches. The third type of action, {\fontfamily{qcr}\selectfont Act.Type3}, considers redirecting the low-priority contents' requests between edge caches. This type of action affects the values of $\mathbb{Z}^{c_l}(t)$ and we denote it by a binary matrix $a_{c_l}^\mathbb{Z}(t)$ of size $N\times N$. Thus, three types of action can be recognized, {\fontfamily{qcr}\selectfont Act.Type1}, {\fontfamily{qcr}\selectfont Act.Type2} and {\fontfamily{qcr}\selectfont Act.Type3}, implemented by $a_{c_l}^\mathbb{Y}(t)$, $a_{c_h}^\mathbb{X}(t)$, and $a_{c_l}^\mathbb{Z}(t)$ respectively, each indicating the possible changes in the corresponding state vectors' values. The overall action $a_t$ at time $t$ is then summarized by
\begin{equation}\label{actions}
     a_{t}=\{<a_{c_l}^\mathbb{Y}(t), a_{c_h}^\mathbb{X}(t), a_{c_l}^\mathbb{Z}(t)> / c_h \in C^{High}(t),c_l \in C^{Low}(t)\}.
\end{equation}

\subsection{Reward Function}
Upon performing an action, the agent needs an immediate feedback to assess the short-term quality of the performed action. This feedback is quantified by the value of a reward function. For the design of this function, we utilized the cost model explained in Section \ref{sec:optimization} (Eqs. \ref{Cmig}, \ref{Caccess} and \ref{Cdeliv}). Our reward function $\mathbb{R}(s_t,a_t)$ is given by
\begin{equation}\label{reward}
     \mathbb{R}(s_t,a_t)= -\left( C_{M}(t) + C_{A}(t) + C_{D}(t) \right), 
\end{equation}
which is defined based on the aggregation of the migration cost $C_{M}(t)$, the low-priority contents' access delay $C_{A}(t)$, and the high-priority contents' download cost $C_{D}(t)$ at the current time slot. Note that the reward function computed by (\ref{reward}) only quantifies the short-term affect of the performed action $a_{t}$ as an immediate feedback.

\begin{figure*}[!t]
\centering
   \includegraphics[width=1\textwidth]{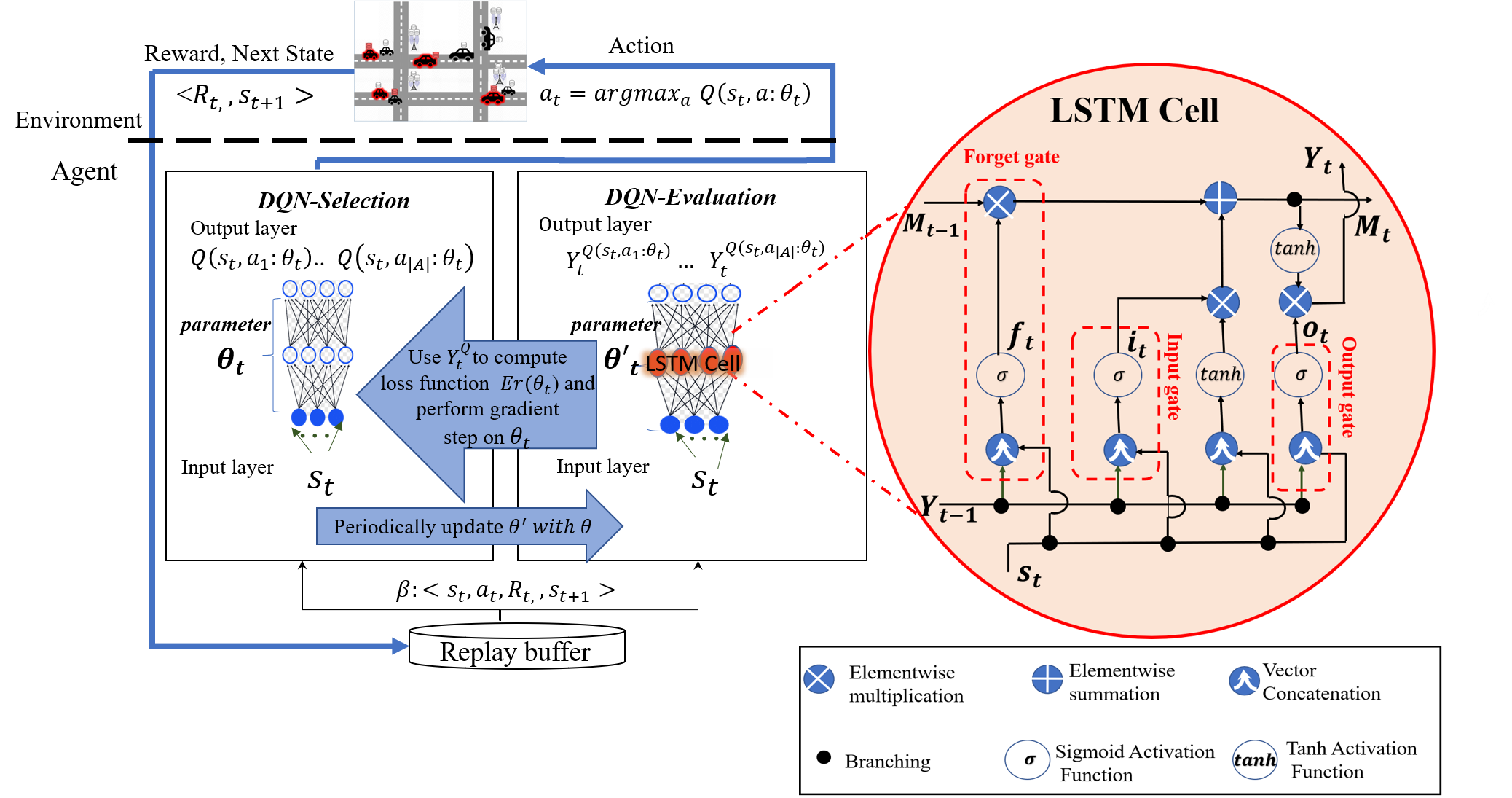}
  \caption{A schematic view of the agent and its interactions with the environment, including the structure of our deployed LSTM cell.}
  \label{Interaction}
 \end{figure*}

\subsection{Design of the deep RL agent}
Unlike non-learning approaches, the RL agent automatically learns the ever-changing environment and updates its decisions through its interactions. Figure~\ref{Interaction} illustrates a schematic view of our agent and these interactions. We will refer to Fig.~\ref{Interaction} and Algorithm~1 as we explain the theoretical steps of our work. Specifically, our approach is based on Q-learning, one of the most widely used RL strategies~\cite{c19}. Q-learning works by successively updating the evaluation of the long term quality (the Q value) of actions at each state. It is a simple way for an agent to learn how to act optimally \cite{c19}. We note, however, that classic Q-learning is limited to tasks with a small number of states and actions~\cite{c20}. Moreover, in Q-learning algorithm, all the states should be met and all the actions should be experienced. Those restrictions are impractical in our problem, as it deals with an environment that is extremely complex and dynamic and its states are large and vary rapidly over time. The only way to learn anything in these types of dynamic situations (where we have dynamic state space) is to generalize from previously experienced states to new states~\cite{c20}. The required generalization is often called function approximation~\cite{c20}. In this study, to approximate the Q values for unmet states/actions, we use a deep neural network (DNN)-based approach, which relies on nonlinear gradient-descent function approximation~\cite{c20}. This approach eliminates the need for visiting all the state/action pairs to compute the Q values. First proposed in \cite{c32}, this revival hybrid approach is now widely used in different domains under the so-called deep reinforcement learning (DRL) or deep Q-learning (DQL) method \cite{c12,c13,c14,c15,c16,c17,c18}.\\

In this work, since our problem concerns a sequential decision-making process, we exploit an advanced version of DQL, a double deep Q-network (DDQN) with LSTM memory cells. In the rest of this section, we first explain the motivation for choosing this specific Q network architecture, and then in Section~\ref{sec:DDQN with LSTM cells}, we discuss the limitations of conventional recurrent neural networks and explain how LSTM memory cells can overcome those limitations, ending with our DDQN algorithm for content migration.

\subsection{DDQN}\label{DDQN}
Q-Learning is a model-free reinforcement algorithm to estimate Q values for state-action pairs. The Q value of a state-action pair can be interpreted as an expected discounted reward accumulated over a long time period. As an example in a given state $s \in S $ ($S$ being the set of all the states) with two possible actions $ a_{1}, a_{2} \in A $ ($A$ being the set of all the possible actions), if $Q(s,a_{1})>Q(s,a_{2})$, then choosing $a_{1}$ over $a_{2}$ will result in a higher accumulated reward over a long term. The detailed mathematical explanation can be found within the well-known Bellman equation~\cite{c19}. The Q-Learning algorithm starts by initializing the Q values for all state-action pairs by setting them to zero. Next, it  recursively computes and updates the Q value of a given pair as follows:
\begin{equation}\label{Q_update}
\begin{aligned}
         Q^{new}(s,a) = &Q^{old}(s,a) \\&+ \alpha \cdot [R + \gamma \cdot \max_{a}Q(s', a) - Q^{old}(s,a)],
\end{aligned}
\end{equation}
where $R$ is the reward of performing action $a \in A$ in state $s\in S$, $s'\in S$ is the next state, $\alpha \in [0,1]$ denotes the learning rate, and $\gamma \in [0,1]$ is the discounting rate. The Q update continues until all the states are met and all the actions have been experienced. At this point, the final Q value, $Q^*(s,a)$, determines the best action $a^* \in A$ at a given state as follows:
$$a^* = \text{argmax}_aQ^*(s,a).$$

It is important to note that in our considered problem, there is no final state, especially given that the states vary over time (mobile caches move and new high-priority contents arrive). Therefore, an approximation method is required for the Q values of unmet states~\cite{c20}. In a deep Q network (DQN), a multi-layered neural network is utilized to estimate the Q values. At state $s_t$, the learning agent takes action $a_t$ based on policy $\varepsilon$, which is initially purely random and gradually improves as the agent becomes more experienced. Let us denote the reward and the resulting state as $R_t$ and $s_{t+1}$, respectively. The tuple $ e_t = <s_t, a_t,R_t, s_{t+1}>$ represents the experience of the agent at time $t$ stored in a buffer called the experience replay buffer. Periodically, the samples of the agent's experience will be drawn randomly to form the learning batches. These learning batches are then used to feed the DQN and update the estimated Q values.

For a given state-action pair $<s_t, a_t>$, $Q(s_t, a_t;\mathbf{\theta_{t}})$ is the DQN current estimation of the Q value. Here, $\mathbf{\theta_{t}}$ is the parameter of the Q network at time $t$. The gradient descent update rule for the parameter $\mathbf{\theta_{t}}$ will be applied as follows: 
\begin{equation}\label{gradient}
       \theta_{t+1} = \theta_{t} + \alpha(Y_{t}^Q- Q(s_t, a_t;\mathbf{\theta_{t}})) \cdot \nabla_{\mathbf{\theta_{t}}} Q(s_t, a_t;\mathbf{\theta_{t}}),
\end{equation}
where $\alpha$ is the gradient step size and $Y_{t}^Q$ denotes the target Q value with current parameter $\mathbf{\theta_{t}}$, which is calculated by 
\begin{equation}\label{target}
     Y_{t}^Q = R_t + \gamma \cdot \max_{a}Q(s_{t+1}, a;\mathbf{\theta_{t}}).
\end{equation}
With the update rule (\ref{gradient}), the parameter $\mathbf{\theta_{t}}$ of the DQN will be tuned so that $Q(s_t, a_t;\mathbf{\theta_{t}})$ moves towards $Y_{t}^Q$ with step size $\alpha$. Note, however, that in doing so, $Y_{t}^Q$ itself is computed by the maximum value of $Q(s_{t+1}, a;\mathbf{\theta_{t}})$, as shown in (\ref{target}). This loop, in turn, will cause an over-optimistic and unstable approximation of the Q values, which can degrade the accuracy of the results~\cite{c33}. This can be avoided using a technique first proposed by Van Hasselt~\cite{c33}, where two DQNs are trained in parallel.
The first DQN, $Q^{Select}(s,a;\mathbf{\theta_{t}})$, referred to as \textit{\text{DQN}-selection}, with parameter $\mathbf{\theta_{t}}$ is used for the selection of actions, whereas the second DQN, $Q^{Eval}(s,a;\mathbf{\theta'_{t}})$, referred to as \textit{\text{DQN}-evaluation}, with parameter $\mathbf{\theta'_{t}}$ is trained for the evaluation of the actions. With these settings, instead of Eq.~(\ref{target}), the target Q value for the DDQN will be computed as follows:
%
%
\begin{equation}\label{DDQN_target}
\begin{aligned}
             Y^Q_{t} =  R_t + \gamma \cdot Q^{Eval}(s_{t+1}, \text{argmax}_a Q^{Select}(s_{t+1},a;\mathbf{\theta_{t}});\mathbf{\theta'_{t}}).
\end{aligned}
\end{equation}

Accordingly, the error function $Er(\mathbf{\theta_{t}})$ of DDQN at time $t$ is given by:
\begin{equation}\label{Error}
\begin{aligned}
        Er(\mathbf{\theta_{t}}) = \frac{1}{2} [Y^Q_{t} - Q^{Select}(s_t,a;\mathbf{\theta_{t}})]^{2}.
\end{aligned}
\end{equation}

After each forward pass, $\text{Er}(\mathbf{\theta_{t}})$ will be recalculated. Following the back propagation procedure and the derivation chain rule, the contribution of each DDQN parameter to the error will be obtained. The gradient descent update rule given by Eq.~(\ref{gradient}) uses this calculated value to update the parameters. Periodically, the values of $\mathbf{\theta_{t}}$ will be copied to $\mathbf{\theta'_{t}}$. After sufficient training, the parameters will be tuned such that the error value becomes quite small. For illustration, we depict the interactions of two networks $Q^{Select}$ and $Q^{Eval}$ in Fig.~\ref{Interaction}. 
The decoupling of the selection and evaluation Q networks in the learning process has proven to be successful for reducing over-optimism and to therefore produce more stable and reliable learning results \cite{c31}.
\subsection{DDQN with LSTM cells}\label{sec:DDQN with LSTM cells}
Here, we explain the theoretical advantage of using LSTM memory cells in our designed DRL agent. In Section~\ref{sssec:num3}, we will demonstrate this advantage with an analytical discussion on the concrete behavior of our agent. 
\par The estimation of Q values becomes more accurate as the agent becomes more experienced, given that the current estimation for Q values is based on the agent's experienced states, actions, and rewards. However, as the DDQN continues to learn, the impact of some important experiences in the distant past could be replaced by more recent experiences. This problem, which is also referred to as the vanishing gradient \cite{c32,c33,c35}, is a well-known obstacle in the learning path of gradient-based approaches such as RNN~\cite{c34}. The vanishing gradient makes the learning process time-consuming and may lead to inaccurate results~\cite{c10}. In the following, we discuss the theoretical aspects of the vanishing gradient and then explain the high-level solution that LSTM memory cells provide in this regard. 
\par Consider the gradient update rule in a DNN given by Eq.~(\ref{gradient}). After passing many gradient update steps and when $t$ becomes large enough, the error and the gradient term $(Y_{t}^Q- Q(s_t, a_t;\theta_t)) \cdot \nabla_{\theta_t} \cdot Q(s_t, a_t;\theta_t))$ 
becomes so small that the values of $\theta_t$ do not change significantly. Insufficient decaying error backflows to the initial layers of the neural network, thus hampering the learning process~\cite{c32}. To avoid this issue, the authors of ~\cite{c32} have suggested using long short-term memory (LSTM) cells, which are deployed in the hidden layers of the given DNN to ensure the flow of decaying error in the backpropagation process in later learning steps, thereby allowing the learning process to continue. It is worth noting that the LSTM architecture\footnote{For a detailed discussion of LSTM computational components, the interested reader is referred to~\cite{c21}, which includes a systematic study of various LSTM architectures.} is now widely used in many DNN applications \cite{c35,c36,c37} and has been proven to outperform the simple feedforward DNNs \cite{c34}. 
Figure \ref{Interaction} depicts the structure of our deployed LSTM cell. The LSTM cell is comprised of three inputs, $\mathbf{M_{t-1}, Y_{t-1}}$, and $\mathbf{s_{t}}$, which are the previous memory state of the cell, the previous output of the cell (i.e., the previous predicted value), and the current input of the network, respectively. The two inputs $\mathbf{M_{t-1}}$ and $\mathbf{Y_{t-1}}$ of the cell are initialized to be all zeros at time $t=0$. The LSTM cell outputs two vector values, $\mathbf{Y_{t}}$, and $\mathbf{M_{t}}$ which are the current output (i.e. predicted value), and the current memory state of the cell, respectively. As shown in Fig.~\ref{Interaction}, an LSTM cell consists of three gates: ($i$)~forget, ($ii$)~input, and ($iii$)~output, each containing a sigmoid activation function denoted by $\sigma(x) = (1+ e^{-x})^{-1}$. The output of the sigmoid functions of the forget, input, and output gates are $f_t$, $i_t$, and $o_t$, respectively. Each of these activation functions has it own weights and bias as follows: $\mathbf{W_f}$ and $\mathbf{b_f}$ for the forget gate, $\mathbf{W_i}$ and $\mathbf{b_i}$ for the input gate, and $\mathbf{W_o}$ and $\mathbf{b_o}$ for the output gate. All these parameters are randomly initialized at the beginning. With these settings, the forward pass formulas of an LSTM cell are as follows:
$$\mathbf{f_t} = \sigma (\mathbf{W_f}[\mathbf{Y_{t-1}},\mathbf{s_t}]+ \mathbf{b_f}),$$
$$\mathbf{i_t} = \sigma (\mathbf{W_i}[\mathbf{Y_{t-1}},\mathbf{s_t}]+ \mathbf{b_i}),$$
$$\mathbf{o_t} = \sigma (\mathbf{W_o}[\mathbf{Y_{t-1}},\mathbf{s_t}]+ \mathbf{b_o}),$$
$$\mathbf{M_{t}} = \mathbf{M_{t-1}} \otimes \mathbf{f_t} \oplus (\mathbf{i_t} \otimes \tanh([\mathbf{Y_{t-1}},\mathbf{s_t}])),$$
$$\mathbf{Y_t} = \tanh(\mathbf{M_{t}}) \otimes \mathbf{o_t},$$
where $[\mathbf{Y_{t-1}},\mathbf{s_t}]$ is the concatenation of vectors $\mathbf{Y_{t-1}}$ and $\mathbf{s_{t}}$, while the element-wise multiplication and summation are denoted as $\otimes$ and $\oplus$, respectively, and $\tanh$ is the hyperbolic tangent function.
During training, the cell parameters $\mathbf{W_i}$, $\mathbf{b_i}$, $\mathbf{W_o}$, $\mathbf{b_o}$, $\mathbf{W_f}$, and $\mathbf{b_f}$ are tuned using the back propagation and stochastic gradient descent update rules explained in Section~\ref{DDQN}. Note in our proposed algorithm, the LSTM cell is embedded in the hidden layers of the $\text{DQN}-Evaluation$, as shown in Fig.~\ref{Interaction}.
\par Algorithm \ref{euclid} illustrates the main steps of our double deep Q-Learning algorithm used for solving our content migration problem.
The algorithm starts with observing the initial state, $s_1$. A series of iterations are then followed while the algorithm switches between exploration and exploitation phases. Parameters $\epsilon$ (exploration rate), and $\lambda$ (a random value in range $[0,1]$) are used to control these phases. A random action and its type are selected in the exploration phase (see lines 10-23), while in the exploitation phase, $\text{DQN}-selection$ will determine the action (see line 25). The action, reward, and next state are then collected and stored in buffer $\mathcal{D}$ (line 29). A batch of experiences is then randomly retrieved from $\mathcal{D}$ (line 30). The target value of $\text{DQN}-selection$ (i.e., $Y_t^Q$) can be computed by the use of $\text{DQN}-evaluation$ (line 31). This target value will be used for computing  the error function $Er(\theta_t)$, which is the average error of all samples of $\beta$ (line 32). The parameters of $\text{DQN}-selection$ will be updated by performing a gradient descent step on $Er(\theta_t)$ with respect to $\theta_t$ (line 33). Finally, every $\bar\tau$ steps the parameters of the $\text{DQN}-selection$ are copied to $\text{DQN}-evaluation$ (line 34).
\begin{algorithm}[!t]
\scriptsize
	\caption{DDQN-based Content Migration}\label{euclid} 
	\begin{algorithmic}[1]
	\State Initialize the \textbf{\textit{\text{DQN}-selection}} network with $\theta_t = \theta_0$
    \State Initialize the \textbf{\textit{\text{DQN}-evaluation}} network with $\theta^\prime_t = \theta^\prime_0$ 
    \State Initialize two vectors $a_{c_l}^\mathbb{Y}$, $a_{c_h}^\mathbb{X}$ of size $N$ and a matrix $a_{c_l}^\mathbb{Z}$ of size $N \times N$.
    \State Initialize replay buffer $\mathcal{D}$ 

		\For {episode $k = 1$ to $K$}
		    \State Observe the current content placement and delivery state of the environment and construct $s_1$ as 
		    Eq.~(\ref{S}).
		    \For{time slot $t = 1$ to $T$}
		        \State Generate a random number $\lambda\in[0,1]$
		        \If{$(\lambda < \varepsilon)$}
		            \Comment{Exploration phase}
		            \State reset $a_{c_l}^\mathbb{Y}(t)$,  $a_{c_h}^\mathbb{X}(t)$, and $a_{c_l}^\mathbb{Z}(t)$ with zero values.
		            \State $Action\_Types = [${\fontfamily{qcr}\selectfont Act.Type1}, {\fontfamily{qcr}\selectfont Act.Type2}, {\fontfamily{qcr}\selectfont Act.Type3}$]$
		            \State Choose a random integer $i \in \{0, 1, 2\}$
		            \If{$( Action\_Types[i] ==$ {\fontfamily{qcr}\selectfont Act.Type1}) }
		                \State Choose a random integer number $j \in \{0,...,N-1\}$
		                \State Assign a value of 1 to the $j^{th}$ element of $a_{c_l}^\mathbb{Y}(t)$.
		             \ElsIf{$( Action\_Types[i] == $ {\fontfamily{qcr}\selectfont Act.Type2}) }
		                \State Choose a random integer number $l \in \{0,...,N-1\}$
		                \State Assign a value of 1 to the $l^{th}$ element of $a_{c_h}^\mathbb{X}(t)$.	
		             \ElsIf{$( Action\_Types[i] == $ {\fontfamily{qcr}\selectfont Act.Type3}) }
		                \State Choose two random integer numbers $m,n \in \{0,...,N-1\}$
		                \State Assign a value of 1 to the $(m,n)^{th}$ element of $a_{c_l}^\mathbb{Z}(t)$.
		                
		             \EndIf
		             \State construct $a_t$ as
		              Eq.~(\ref{actions}).
                \Else
                    \Comment{Exploitation phase}
                    \State Choose $a_t$ which maximizes $Q^{Select}(s_t, a_t;\theta_t)$
                \EndIf
			    \State Execute $a_t$ 
			    \State Observe the next state $s_{t+1}$ and reward $R_t$
				\State Store $<s_t, a_t,R_t, s_{t+1}>$ in replay buffer $\mathcal{D}$
				\State Get a batch of experiences $\beta$ randomly sampled from $\mathcal{D}$
				\State Set target $Q$ value $Y_t^Q$ : \newline 
				\hspace*{10mm} $Y_t^Q = R_t + \gamma \cdot Q^{Eval}(s_{t+1},\text{argmax}_a Q^{Select}(s_{t+1},a;\theta_t);\theta^\prime_t)$
				\State Set the average error function $Er(\theta_t)$: 
				\newline 
				\hspace*{10mm} $Er(\theta_t) = \frac{1}{\beta} \sum^\beta_{i=1}{(Y_i^Q-Q^{Select}(s_i, a_i;\theta_t))}^2$
				\State Update $\theta_t$ with the gradient descent update rule:\newline \hspace*{9mm}  $\theta_{t+1} = \theta_t - \alpha \frac{\partial Er(\theta_t)}{\partial \theta_t}$
				\State Every $\bar\tau$ iterations set $\theta_t^\prime = \theta_t$
			\EndFor
		\EndFor
	\end{algorithmic} 
	
\end{algorithm}
\section{SIMULATION RESULTS AND DISCUSSIONS}\label{sec:evaluation}
To ensure that our simulated evaluations are conducted based on realistic scenarios, we used the SUMO (Simulation for Urban MObility) simulator~\cite{c38}. This platform is widely used for evaluating new approaches in vehicular networks. We rely on this simulator to provide realistic vehicle traces, which are required as part of the inputs for the training and testing of the proposed deep RL agent. 
As for the DDQN, we used TensorFlow 1.6.0~\cite{c39}, Google's open-source machine learning library. In particular, we utilize ``{\fontfamily{qcr}\selectfont tf.contrib.rnn.LSTMCell}'' and ``{\fontfamily{qcr}\selectfont keras.models}'' classes to instantiate the two four-layer DNNs, \textit{\text{DQN}-selection} and \textit{\text{DQN}-Evaluation}, with LSTM cells in hidden layers of the latter DNN. To assure convergence, we rely on the Keras class ``{\fontfamily{qcr}\selectfont ReduceLRONplatue}" to automatically update the learning rate. All the simulation tests were conducted on a machine with 2.67 GHz Intel Xeon CPU E5640 and 32~GB of memory.
 
 \subsection{Simulation Settings}\label{Ssec:Simulation Settings}
In our simulations, we considered an $n\times n$ bidirectional road grid environment \cite{c28} where each grid cell covers an area of 0.25~km$^2$. The number of grid cells, mobile caches and fixed caches are specified in each evaluation scenario.
In this grid structure, mobile caches move with an average velocity of 30~km/hour and with the parameters $\mu_{\mathscr{S}} = 0.5$, $\mu_{\mathsf{L}} = 0.25$ and $\mu_{\mathscr{R}} = 0.25$, set according to the Manhattan model, the most popular model for mobility in urban areas~\cite{c28}.
Each fixed cache has a capacity of 1.5~GB memory and 4 processing units. Each mobile cache was provided with a 700~MB memory and 1~processing unit. We assume that the fixed edge caches can handle up to 20 requests at a time and that each mobile cache can handle a maximum of 5 requests simultaneously. While the fixed edge caches have a 60-meter diameter circular coverage, the mobile caches can cover a circular area of only 10 meters in diameter. In addition, 40 low-priority contents of various sizes from 50 to 120~MB are randomly placed so that 8 of the mobile caches are fully occupied. 

At the beginning, we assume that 5 of these full mobile caches are targeted caches. In the simulation scenario, high-priority contents arrive according to a Poisson process with an average arrival rate of 5 contents per time unit. The sizes of these contents are similar to the sizes of the low-priority contents' ranges: from 50 to 120~MB. We set the delay for transmitting 1~MB of content (high or low priority) from cloud servers to an edge cache node (fixed or mobile), and the delay from each cache node to the users of its coverage as 0.5 and 0.2 milliseconds \cite{c49}, respectively. The average cost of the power consumption required by edge caches to upload and download 1 MB is set to 2 units of currency, while the average cost of transferring 1 MB between the edge caches for one hop is set at 3 units. The costs of each second of delay in accessing 1 MB of low- and high-priority content are set to 5 and 10 units of currency, respectively (timely access to high-priority contents is critical and so its delay costs twice as much). We consider the same parameter settings for all the following simulated scenarios unless otherwise stated. To set up the learning process, the actions selected in the initial 700 time slots are totally random and are used initially to fill the experience replay buffer in order to start the learning process. In each time slot (for $t>700$), 30 samples of experience are extracted from the replay buffer to form the learning batch. 
\vspace{-0.1cm}
\subsection{Comparison with the optimal solution}
\begin{figure}[ht]
\centering
\includegraphics[width=0.5\textwidth]{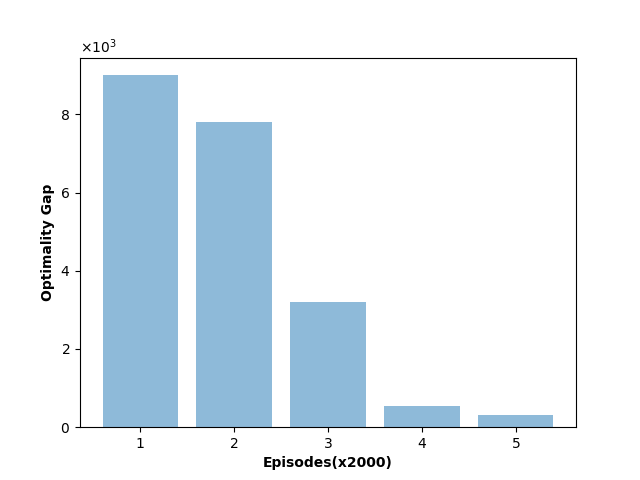}
\caption{Optimality gap of our proposed DRLCM method vs. episode.}
\label{dif}
\end{figure}

Finding the optimal solution for the content migration problem described in Section~\ref{sec:objective_function} is very time-consuming. To evaluate the performance of our proposed DRLCM algorithm against that of the optimal solution, we consider a small-scale scenario with only two mobile and two fixed edge caches. The road structure in this scenario is a $2\times2$ bidirectional road grid,with the other parameters the same as the scenario explained in  \ref{Ssec:Simulation Settings}. We let our learning method collect experiences about the environment while learning for a maximum of 10,000 episodes. We pause the simulation scenario every 2,000 episodes and perform an exhaustive search to find the optimal solution considering the positioning of mobile caches and content arrivals at that time. Figure \ref{dif} shows the absolute difference between the optimal solution and the DRLCM at every 2,000th episode. While the method is still in its exploration phase, the gap is considerably higher for the initial episodes. However, it decreases significantly as the episodes pass. At the end of 10,000 episodes, the DRLCM managed to decrease this gap by more than 97\%. An important consideration here is that it takes more than 1 hour for each exhaustive search to find the optimal solution in this small scale scenario and that is only one snapshot of the whole system. Clearly it is not possible to conduct  exhaustive searches each time mobile edge caches change their position.

\subsection{Performance comparison with existing deep learning methods}\label{sssec:num3}
We investigate the convergence performance of our proposed DRLCM with two other deep learning approaches, namely, SRLCM, a simplified Q-Learning version with Double RNNs and no LSTM cells \cite{c17}, and TRLCM, a learning method with a single RNN and no LSTM cells ~\cite{c32}, i.e., a traditional deep Q-Learning method. While the learning structure of the SRLCM method has been widely used in many recent studies \cite{c17,c18,c31}, TRLCM represents a classical version of the deep Q-Learning approach \cite{c32}. Note that apart from the differences in their learning structures, the three methods: DRLCM, SRLCM, and TRLCM use the same experience reply technique and the same learning batch size (30 samples in each batch), as well as the same automated learning rate updates. The evaluation scenario consists of 12 fixed  and 20 mobile edge caches in a $5\times 5$ road grid environment.

\begin{figure}[!t]
\begin{center}
\includegraphics[width=.5\textwidth]{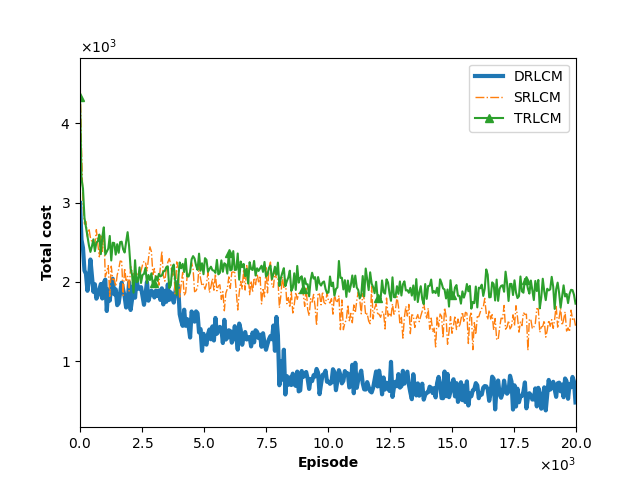}
\caption{Total cost vs. episode for different content migration-based deep learning methods and our proposed DRLCM method.}
 \label{Converge}
\end{center}
\end{figure}

Figure~\ref{Converge} depicts the total cost (in unit of currency) vs. episodes for different methods. According to Fig.~\ref{Converge}, all three deep learning-based methods perform closely for the first episodes. This is mainly due to the fact that at the beginning, there is no knowledge about the environment and so all the methods choose somewhat random actions. However, due to their different learning structures, they converge to different values. The policy learning of TRLCM seems to stop soon after completing 7,500 episodes, whereas the total costs achieved by the SRLCM and DRLCM methods keep decreasing. Finally, around the 15,000 episode, the SRLCM method reaches a cost value of $~$1,500 and levels out. In contrast, our proposed DRLCM method continuously decreases the total cost as the number of episodes increases. Clearly, our proposed DRLCM method outperforms the other two deep learning methods. This high performance is attributed to the use of LSTM memory cells, which allow the DRLCM agent to remember the most valuable experiences that it had in its past observations. To see this more clearly, let us consider the highlighted points in Fig.~\ref{Converge}. At these points (around episodes 5,000 and 7,500), random actions of the agent resulted in sudden decreases in cost. More precisely, while the algorithm was in its exploration phase, two random placement of contents contributed to very good results, i.e., low-cost solutions. Since lowering the cost is desired in our problem, we can say around these particular episodes,  the agent had very valuable experiences. Even though the environment will change after these experiences (mobile caches will move and new requests will arrive) we still want the agent to remember them and avoid overriding their effects by more recent experiences. Empowering the DRLCM with LSTM memory cells automatically enabled this objective without any explicit manual control. The LSTM cells' weights and biases determine what to remember and what to forget, and they are tuned in parallel with the Q networks (see Section~\ref{sec:DDQN with LSTM cells}). 

%
%

\begin{figure}[!t]
\begin{center}
\includegraphics[width=.5\textwidth]{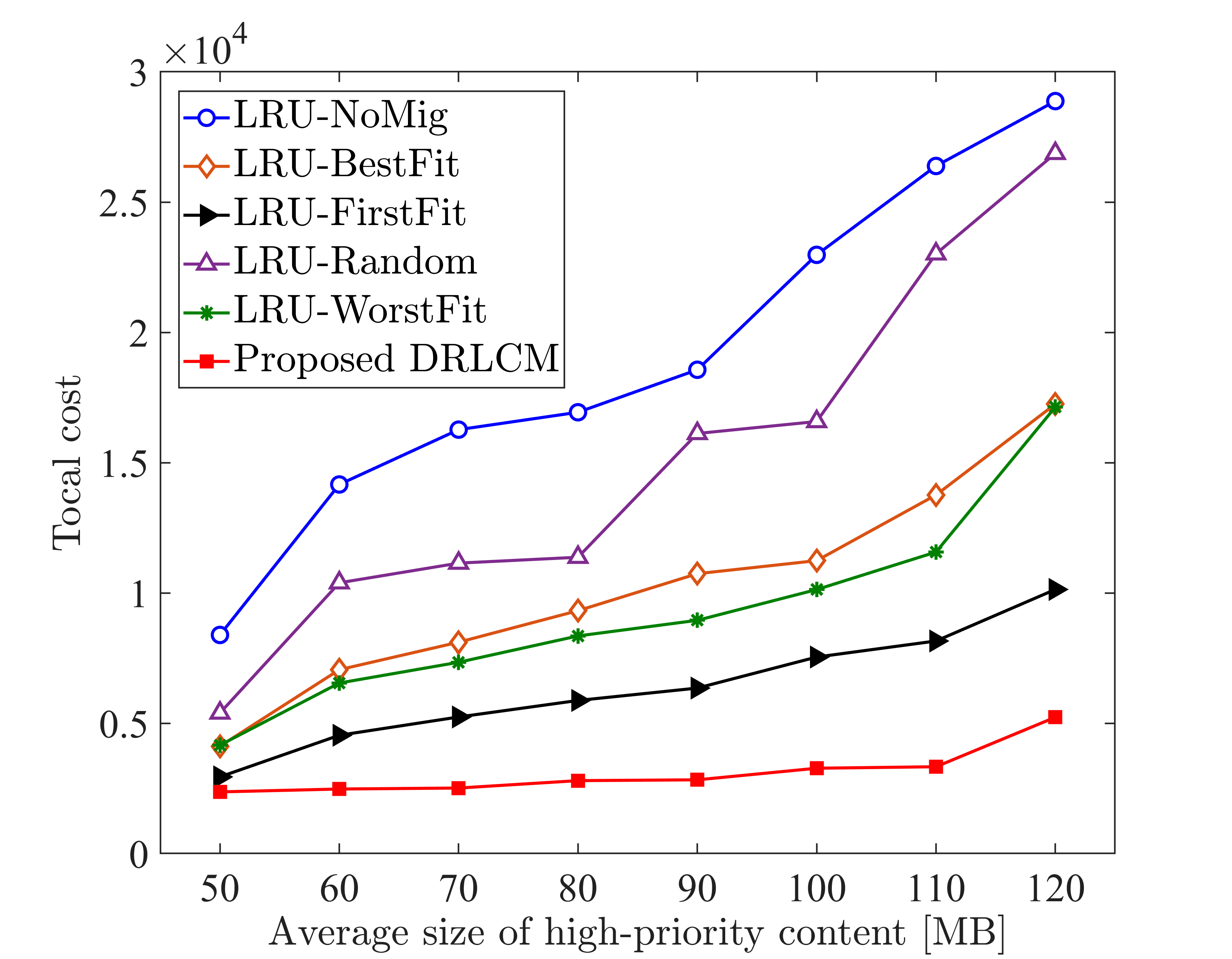}
\caption{Total cost vs. average size of high-priority content for different content eviction strategies and our proposed DRLCM method.}
\label{TotalCost}
\end{center}
\end{figure}

\subsection{Performance comparison with non-learning methods}
\begin{figure}[!t]
\begin{center}
\includegraphics[width=.5\textwidth]{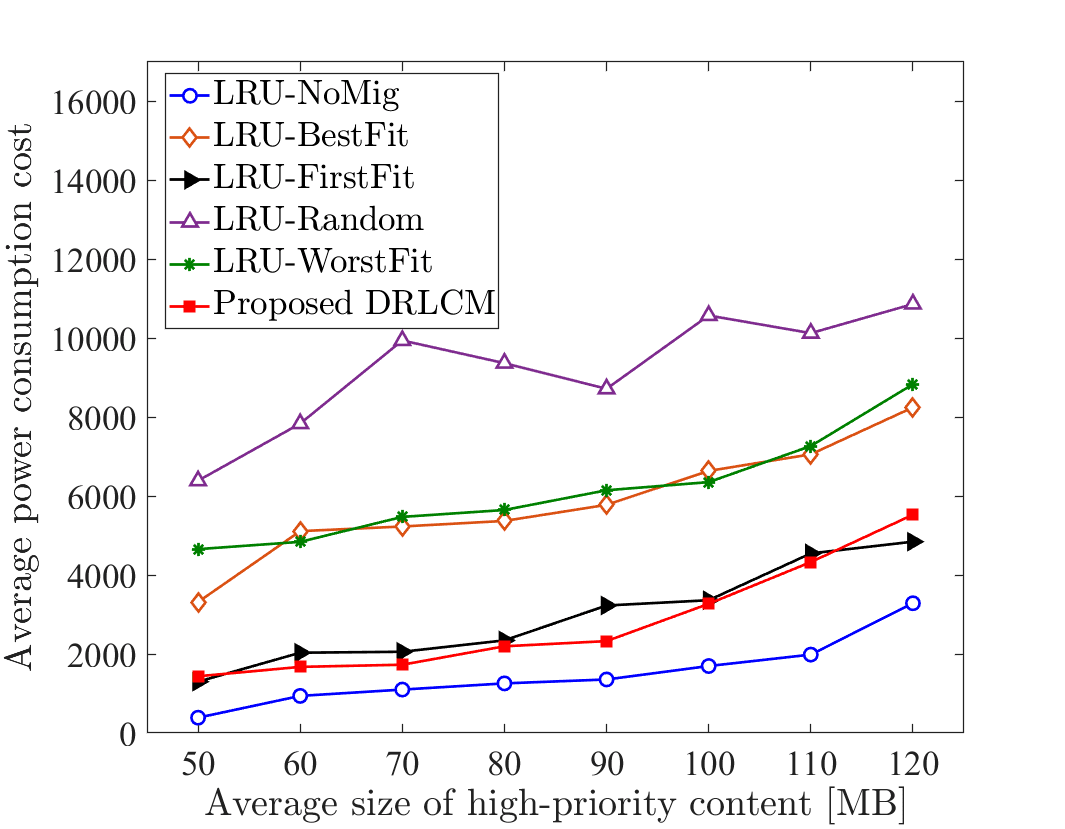}
\caption{Average power consumption cost vs. average size of high-priority contents for different migration methods.}
\label{Power}
\end{center}
\end{figure}

In the next set of evaluation scenarios, we compare the performance of our proposed method with methods based on the least recently used (LRU) eviction strategy~\cite{c10}, which is the most common non-learning cache replacement method. The five LRU-based approaches are explained below:  
\begin{itemize}
\item \textbf{LRU--NoMig:} LRU contents are deleted from the targeted full caches to free up space for the newly arrived high-priority contents.
\item \textbf{LRU--FirstFit:} LRU contents are migrated from the targeted full caches to the closest edge caches that have enough capacity to store them.
\item \textbf{LRU--BestFit:} LRU contents are migrated from the targeted full caches to the edge caches with the minimum caching capacities that can accommodate the migrated contents.
\item \textbf{LRU--WorstFit:} LRU contents are migrated from the targeted full caches to the edge caches with maximum caching capacities that can accommodate the migrated contents.
\item \textbf{LRU--Random:} LRU contents are migrated from the targeted full caches to the random edge caches with enough space to accommodate the migrated contents. 
\end{itemize}
The total cost vs. the average size of arrived high-priority contents (in MB) is shown in Fig.~\ref{TotalCost}, which helps to compare the performance of the five non-learning methods with that of our proposed DRLCM method. As shown in Fig.~\ref{TotalCost}, when the size of high-priority contents increases from 50 to 120 MB, the cost increases in all methods, which is expected, as all methods try to free up more space to accommodate such high-priority contents. Therefore, a larger amount of content will be migrated/deleted and higher costs will be imposed. Further, we observe from Fig.~\ref{TotalCost} that the costly process of re-downloading the deleted contents imposes the largest cost to the LRU--NoMig method. The cost of the LRU--Nomig is even slightly larger than that of the LRU--Random approach, which randomly migrates LRU contents to the available edge caches instead of deleting them. Also, Fig.~\ref{Power} shows the average power consumption cost (per cache) imposed by uploading and/or downloading contents as well as that incurred by transmitting them between edge caches. The LRU--NoMig approach imposes the smallest amount of power consumption cost since it uploads no content and simply drops the LRU contents. Since it only downloads high-priority contents, the average power consumption cost per cache for the LRU--NoMig method increases almost linearly as the high-priority content size increases. The other methods suggest content uploads and downloads to different edge caches for content migration, with the inevitable associated power costs. The DRLCM and LRU--FirstFit methods operate in a competitive manner and have lower power consumption than all but the LRU--NoMig approach, as they target migration within the shortest hop distances. 

\begin{figure}[!t]
\begin{center}
\includegraphics[width=.5\textwidth]{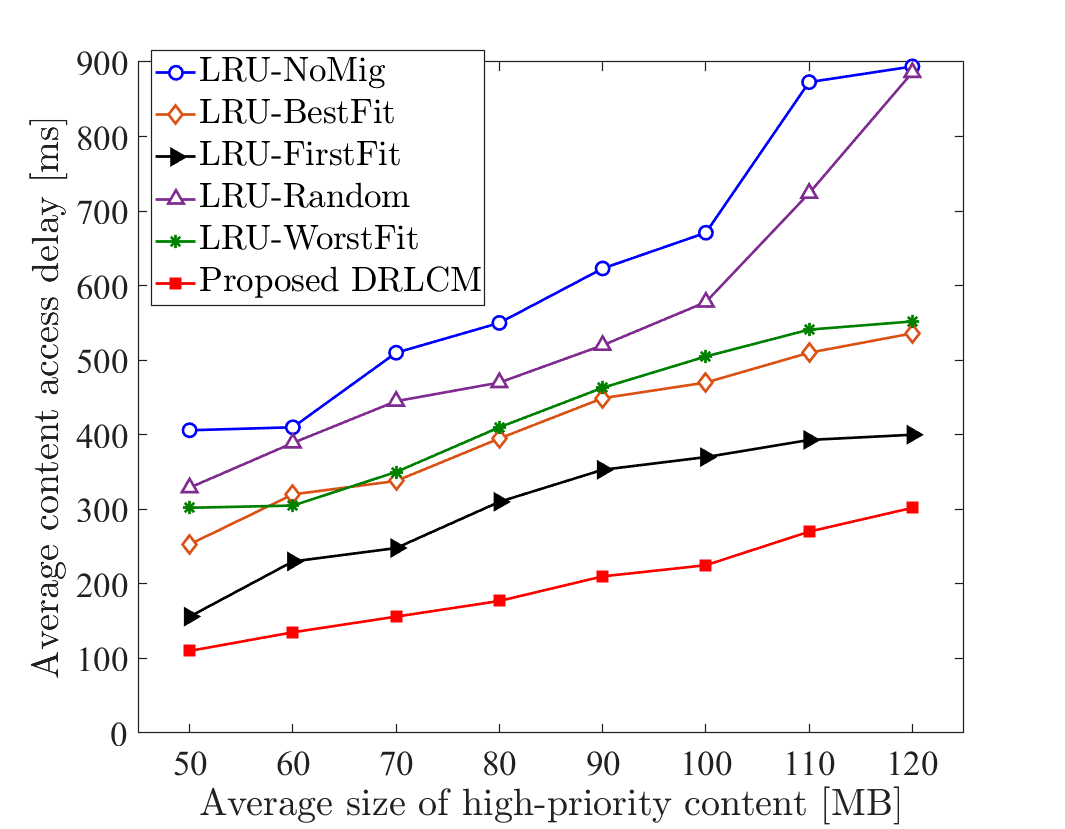}
\caption{Average content access delay vs. average size of high-priority contents for different migration methods.}
\label{Delay}
\end{center}
\end{figure}

Figure~\ref{Delay} illustrates the average content access delay vs. the average size of arrived high-priority contents for different migration methods. The LRU--NoMig approach imposes a high delay, since re-downloading the previously-deleted contents is quite time-consuming. The other LRU--based methods perform marginally better than LRU--NoMig because they manage to keep more content at the edge, instead of complete deletion followed by a cache-miss issue. As shown in Fig.~\ref{Delay}, our proposed DRLCM method outperformed all five LRU-based methods in terms of content access delay, as it directly considers the access delay of the contents in its content migration decision-making process.
%

\begin{figure}[ht]
\centering
\includegraphics[width=0.5\textwidth]{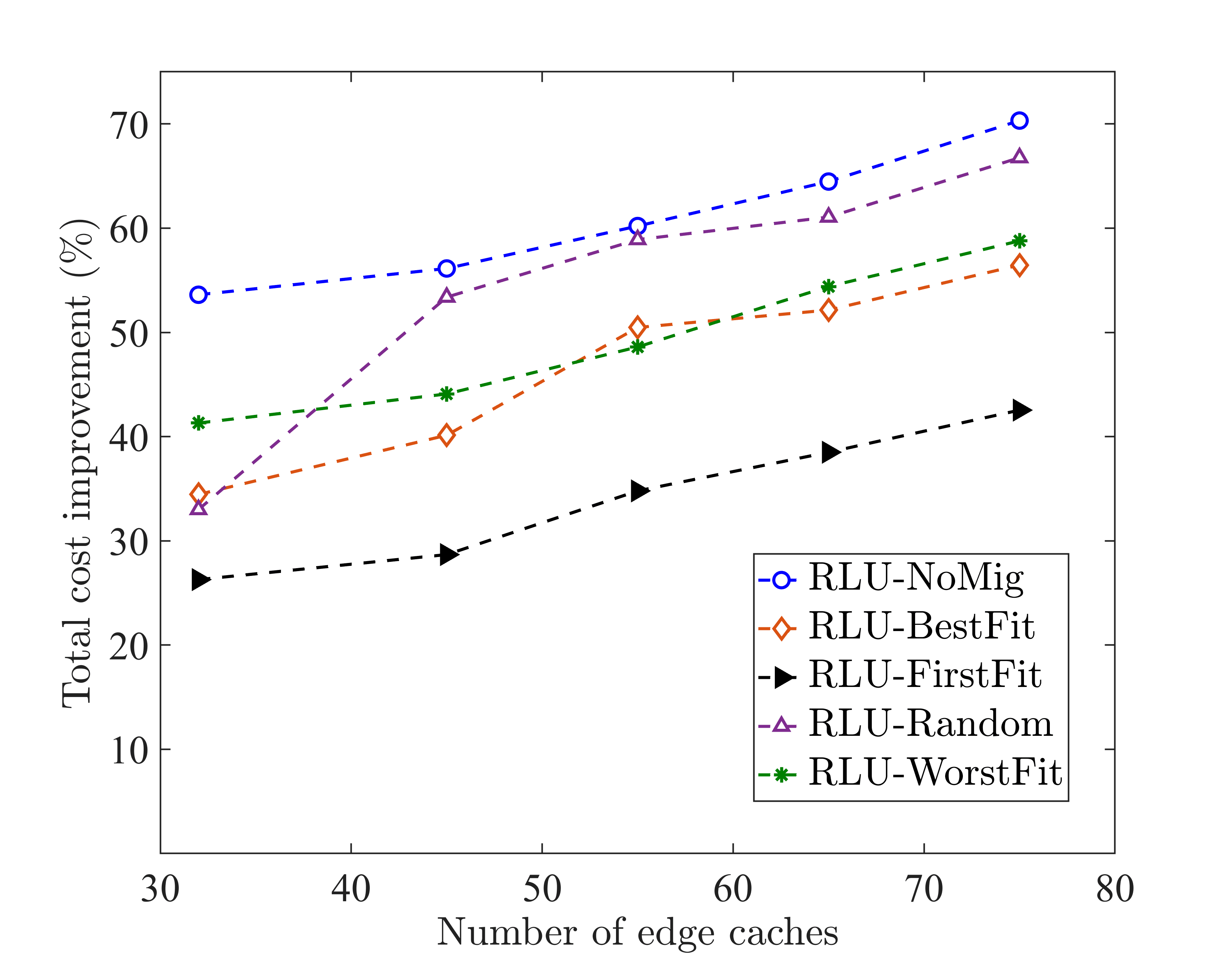}
\caption{Total cost improvement vs. the number of edge caches.}
\label{Improvement}
\end{figure}

 
\subsection{Scalability and cost improvement percentages}
In this evaluation scenario, we increased the total number of edge caches from 32 to 75, and the high-priority content sizes from 50 to 120~MB. The goal was to assess the scalability of our proposed DRLCM method compared to that of the LRU-based approaches. Figure~\ref{Improvement} depicts the improvement of cost (in percentage) vs. the number of edge cashes. It can be inferred from Fig.~\ref{Improvement} that the improvement made by our proposed DRLCM method not only remains for a scaled version of the scenario, but increases up to 70\% in comparison with the LRU-NoMig approach, which does not support any content migration. This observation reveals the value of an appropriate decision to keep the content at the edge instead of performing content deletions. Note that even though increasing the number of edge caches makes the scenario more complex, it ensures that more caching and processing resources become available at the edge. We can also observe that the simple decision-making process deployed by LRU-based methods does not have the potential for the effective utilization of increased resources.

\section{Conclusions and future work}\label{sec:conclusions}
We have proposed a deep reinforcement learning (DRL) content migration technique for a hierarchical edge-based CDN. Based on real life situations, we considered a dynamic and heterogeneous environment consisting of mobile and fixed caches where contents have pre-assigned high and low priorities and developed a use case from a vehicular network to illustrate the motivation of our work. Our proposed method considers the available caching capacity in edge caches so that upon the arrival of high-priority contents, instead of just removing the low-priority contents from full caches, it migrates low-priority contents between edge caches to create enough space to accommodate high-priority contents. We implemented our DRL migration agent with a deep double-Q learner method empowered by LSTM memory cells. The simulation results show up to 70\% in cost improvements compared to the existing methods. As a future research, we aim to extend our work by considering an additional caching layer consisting of drone caches. Even though the use of caches installed on drones provides flexibility for content delivery, their high mobility compounds the complexity of the problem and hence requires further investigations.

\section{Acknowledgements}
This work is partially funded by the CHIST-ERA SCORING project through a Quebec FQRNT grant, and also by the Concordia University HORIZON postdoctoral program.

\bibliographystyle{IEEEtran}
\bibliography{mybibliography}

\begin{IEEEbiography}[{\includegraphics[width=0.9in,height=0.8in]{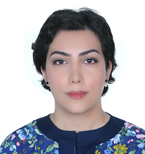}}]{Sepideh Malektaji}
received the B.S. degree in Computer Software Engineering in 2012. In 2015, she completed her M.S. in Computer Architectures. Currently she is a Ph.D. student and a member of Telecommunication Service Engineering Research Lab at Concordia University, Montr\'eal, Canada. In 2016, beginning her Ph.D. program, she received Concordia International Excellence Award. Her current research interests include machine learning application for network resource management and distributed systems, cloud computing, network virtualization, and content delivery networks. She currently serves as a reviewer for many international journals and
conferences. 
\end{IEEEbiography}

\begin{IEEEbiography}[{\includegraphics[width=0.9in,height=0.8in]{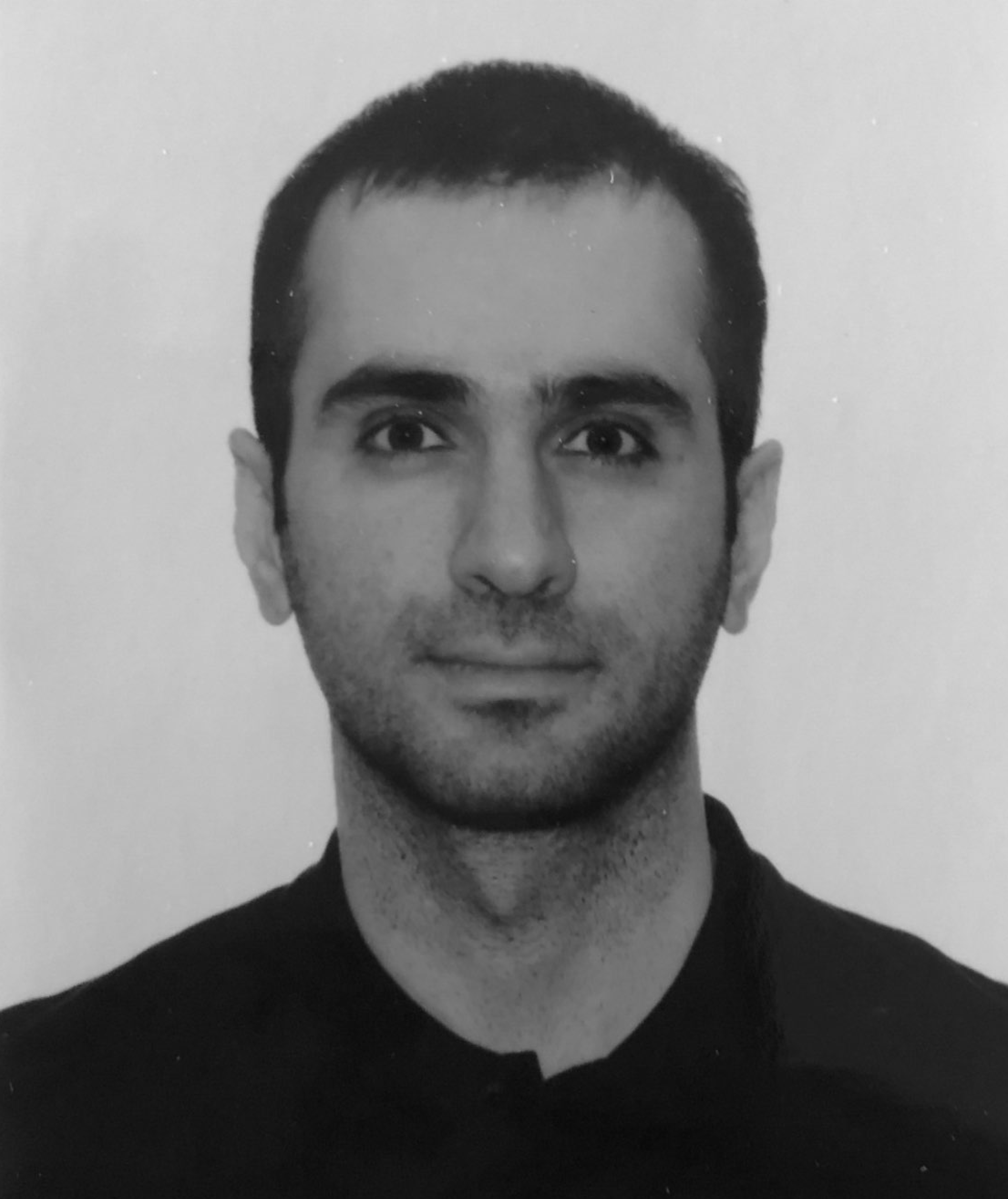}}]{Amin Ebrahimzadeh}
Amin Ebrahimzadeh received the B.Sc. and M.Sc. degrees in electrical engineering from the University of Tabriz, Iran, in 2009 and 2011, respectively, and the Ph.D. degree (Hons.) in telecommunications from the Institut National de la Recherche Scientifique (INRS), Montr\'eal, QC, Canada, in 2019. From 2011 to 2015, he was with the Sahand University of Technology, Tabriz, Iran. He is currently a Horizon postdoctoral fellow with Concordia University, Montr\'eal, QC, Canada. His research interests include 6G networks, Tactile Internet, FiWi networks, and multi-access edge computing. Amin has been awarded the 2019-2020 Best Doctoral Thesis Prize of INRS for his research on the Tactile Internet. He was a recipient of the doctoral research scholarship from the B2X program of Fonds de Recherche du Qu\'ebec-Nature et Technologies (FRQNT). He is the lead author of the book \textit{Toward 6G: A New Era of Convergence} (Wiley-IEEE Press, 2021).
\end{IEEEbiography}

\begin{IEEEbiography}[{\includegraphics[width=0.9in,height=0.8in,clip,keepaspectratio]{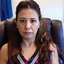}}]{Halima Elbiaze}
holds a Ph.D. in computer science and a M.Sc. in telecommunication systems from the Institut National des T\'el\'ecommunications, Paris, France, and the Université de Versailles in 2002 and 1998. Since 2003, she has been with the Department of Computer Science, Universit\'e du Qu\'ebec à Montr\'eal, QC, Canada, where she is currently an associate professor. In 2005, Dr. Elbiaze received the Canada Foundation for Innovation Award to build her IP over the DWDM network Laboratory. Her research interests include performance evaluation, traffic engineering, cloud computing, wireless networks, and next generation IP networks. She has been awarded many research grants from both public agencies and industry. 
\end{IEEEbiography}
\begin{IEEEbiography}[{\includegraphics[width=0.9in,height=0.8in,clip,keepaspectratio]{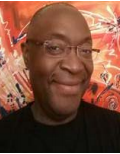}}]{Roch H. Glitho}
(M'88-–SM'97) received the M.Sc. degree in business economics from the University of Grenoble, France, the M.Sc. degree in pure mathematics and the M.Sc. degree in computer science from the University of Geneva, Switzerland, and the Ph.D. (Tech.Dr.) degree in informatics from the Royal Institute of Technology, Stockholm, Sweden. He is currently a Full Professor and a Canada Research Chair with Concordia University. He is also an Adjunct Professor with several other universities, including Telecom SudParis, France, and the University of Western Cape, South Africa. He has worked in industry and has held several senior technical positions (e.g., senior specialist, principal engineer, and expert) with Ericsson, Sweden and Canada. He has also served as an IEEE Distinguished Lecturer, and the Editor-in-Chief of the IEEE Communications Magazine and the IEEE COMMUNICATIONS SURVEYS \& TUTORIALS.
\end{IEEEbiography}

\begin{IEEEbiography}[{\includegraphics[width=0.9in,height=0.8in,clip,keepaspectratio]{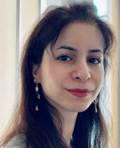}}]{Somayeh Kianpisheh}
received the B.S. degree in software computer engineering from the University of Tehran, Iran, in 2004, and the M.S. and Ph.D. degrees in computer engineering from Tarbiat Modares University, Iran, in 2010 and 2016, respectively. Since 2018, she has been a PostDoctoral Fellow with Concordia University, Canada. She has published several papers in the international journals/conferences and performed reviews for several international journals/conferences. Her research interests include distributed systems particularly resource allocation and performance modeling in 5G, fog/cloud systems, and content driven networks. 
\end{IEEEbiography}

\end{document}